\documentclass[aps,twocolumn,noshowkeys]{revtex4-2}
\usepackage{amsmath}
\usepackage{graphicx}
\usepackage{bm}
\usepackage{hyperref}
\usepackage{xcolor}
\usepackage{multirow}
\usepackage{amssymb}
\usepackage{soul}

\begin{document}
\title{Refined nuclear magnetic dipole moment of rhenium: $^{185}$Re and $^{187}$Re}

\begin{abstract}
The refined values of the magnetic dipole moments of $^{185}$Re and $^{187}$Re nuclei are obtained. For this, we perform a combined relativistic coupled cluster and density functional theory calculation of the shielding constant for the ReO$_4^-$ anion. In this calculation, we explicitly include the effect of the finite nuclear magnetization distribution in the single-particle nuclear model using the Woods-Saxon potential for the valence nucleon. By combining the obtained value of the shielding constant $\sigma=4069(389)$~ppm with the available experimental nuclear magnetic resonance data we obtain the values: $\mu(^{185}{\rm Re})=3.1567(3)(12) \mu_N,  \mu(^{187}{\rm Re})=3.1891(3)(12) \mu_N$,
where the first uncertainty is the experimental one and the second is due to theory. The refined values of magnetic moments are in disagreement with the tabulated values, $\mu(^{185}{\rm Re})=3.1871(3) \mu_N,  \mu(^{187}{\rm Re})=3.2197(3) \mu_N$, which were obtained using the shielding constant value calculated for the atomic cation Re$^{7+}$ rather than the molecular anion. The updated values of the nuclear magnetic moments resolve the disagreement between theoretical predictions of the hyperfine structure of H-like rhenium ions which were based on the tabulated magnetic moment values and available experimental measurements. Using these experimental data we also extract the value of the parameter of nuclear magnetization distribution introduced in [J. Chem. Phys. \textbf{153}, 114114 (2020)], which is required to predict hyperfine structure constants for rhenium compounds.
\end{abstract}

\author{L.V.\ Skripnikov$^{1,2}$, S.D.\ Prosnyak$^{1,2}$}
\affiliation{$^{1}$Petersburg Nuclear Physics Institute named by B.P.\ Konstantinov of National Research Center ``Kurchatov Institute'' (NRC ``Kurchatov Institute'' - PNPI), 1 Orlova roscha, Gatchina, 188300 Leningrad region, Russia}
\affiliation{$^{2}$Saint Petersburg State University, 7/9 Universitetskaya nab., St. Petersburg, 199034 Russia}
\email{skripnikov\_lv@pnpi.nrcki.ru,\\ leonidos239@gmail.com}
\homepage{http://www.qchem.pnpi.spb.ru}

\email{}

\maketitle

\section{Introduction}
Nuclear magnetic dipole moments are of wide interest for many physical problems. They can be used to test predictions of the nuclear theory. They are required as external parameters to predict the hyperfine structure (HFS) of neutral atoms, and molecules. Such data are required to probe the accuracy of calculated electronic wave functions, which are used for calculation of  characteristics of symmetry-violation interactions in atoms~\cite{eEDM_snowmass:2022,Safronova:18,Porsev:2009,ginges2017ground,Skripnikov:2020b,GFreview} and molecules~\cite{KL95,Quiney:98,Titov:06amin,Skripnikov:15b,Skripnikov:15a,Sunaga:16,Fleig:17,Borschevsky:2020,Skripnikov:2020e,Skripnikov:17b}. Such characteristics cannot be directly measured, but they are required to extract the value of the T,P-violating nuclear Schiff and magnetic quadrupole moments, the electron electric dipole moment and other similar effects from the experimental data ~\cite{GFreview,Safronova:18,Skripnikov:14c,Skripnikov:17c,Skripnikov:15c,Skripnikov:17a}. Magnetic dipole moments of stable isotopes can be combined with the experimental and theoretical data on hyperfine structure for stable and short-lived isotopes to obtain magnetic moments of short-lived isotopes~\cite{Persson1998,0954-3899-37-11-113101,Schmidt:2018,barzakh2012hyperfine,Prosnyak:2020,Ginges:2020,Barzakh:2020,Prosnyak:2021}. Magnetic moments are used to predict hyperfine splittings in highly-charged ions, which can be used to test predictions of the bound state quantum electrodynamics~\cite{Shabaev:01a}.

Magnetic dipole moments of stable nuclei can be obtained from nuclear magnetic resonance (NMR) experiments on molecules, though there are suggestions to extract them from precise $g$-factor experiments on highly charged ions~\cite{werth2001g,Quint:2008,Volchkova:2017}. In molecular NMR experiments, one usually obtains so-called uncorrected values of the nuclear magnetic moment. To obtain the intrinsic value of the nuclear magnetic moment, one has to apply a correction on the shielding effect. It is induced by electrons surrounding the nuclei of interest in a given atom or a molecule. Accurate calculation of the shielding constant in molecules containing heavy atoms is rather complicated. Therefore, one often uses shielding corrections calculated for the corresponding atomic ions. Such approach can lead to serious errors~\cite{Skripnikov:18a,Antusek:20,Antusek:18}.

In the present paper we study nuclear magnetic moments for two stable isotopes of rhenium,  $^{185}$Re and $^{187}$Re, both having nuclear spin $I=2.5$. Nuclear magnetic resonance experiments with the aqueous solution of the NaReO$_4$ molecule were carried out in 1951~\cite{Alder:1951}. The tabulated values of the nuclear magnetic moments of $^{185}$Re and $^{187}$Re~\cite{stone2014} are based on those experimental data combined with the shielding constant calculated for the Re$^{7+}$ atomic ion~\cite{Johnson:68,Feiock:1969,Kolb:1982}. In Ref.~\cite{Crespo:1998} it has been noted that such interpretation is not free from possible errors due to neglect of the chemical shift effect, i.e. contribution of molecular environment. The authors of Ref.~\cite{Antusek:20} have used a combination of the nonrelativistic coupled cluster theory and the relativistic density function theory to calculate the molecular shielding constant. Here we perform a precise study of the shielding effect within the relativistic coupled cluster and relativistic density functional theories and show that the completely relativistic treatment allows one to significantly reduce the uncertainty of the shielding constant. In the present paper, we also explore the influence of the finite nuclear magnetization distribution effect in the single-particle approximation on the shielding constant value. Using the refined value of the shielding constant we obtain the updated values of the nuclear magnetic moments of $^{185}$Re and $^{187}$Re. 

The structure of the article is as follows. In Section~\ref{sec:theory}, we give a brief overview of the theory for calculating the shielding constant and the nuclear magnetization distribution effect. In Section~\ref{sectShieldCalc}, we discuss the shielding constant calculation scheme. Section~\ref{sectUncert} provides an analysis of possible uncertainties. These two sections contain some technical details. In Section~\ref{sectNMRmoments}, we derive the nuclear magnetic moments from experimental NMR data and the theoretical value of the shielding constant obtained in sections~\ref{sectShieldCalc} and~\ref{sectUncert}. Section~\ref{sectHlike} analyzes the available experimental data on hyperfine splitting in H-like rhenium ions in various aspects.

The relativistic units ($m$ = $\hbar = c = 1$) and the charge units $\alpha = e^2/(4\pi)$ are used in this paper.

\section{\label{sec:theory}Theory}
One can use the following definition of the shielding tensor corresponding to the nucleus~$j$ in a given molecule:
\begin{equation}
 \label{SHIELDINGDer}
\left.\sigma^j_{a,b}=\frac{\partial^2E}{\partial\mu_{j,a}\partial {\rm B}_b} \right|_{\bm{\mu}_j=0,{\rm \bf{B}}=0}.
\end{equation}
Here $E$ is the energy of the system, $\mu_{j,a}$ is the $a$'th component of the nuclear magnetic moment vector $\bm{\mu}_j$ of $j$'th nucleus, ${\rm B}_b$ is $b$'th component of the uniform external magnetic field vector ${\rm \bf{B}}$. For the interpretation of the molecular NMR experiments, performed in a solution, we need the isotropic part $\sigma$ of the shielding tensor, $\sigma=1/3\sum_a\sigma_{a,a}$. From the nuclear magnetic resonance experiment, it is possible to obtain the uncorrected value $\mu^{\rm uncorr.}$ of the nuclear magnetic dipole moment value, i.e. the value, which is not corrected for the magnetic shielding. The intrinsic value of the magnetic moment $\mu$ can be obtained as:
\begin{equation}
\label{mumuuncorr}
\mu=\mu^{\rm uncorr.} / (1-\sigma).    
\end{equation}

The interaction of electrons in a molecule with an external uniform magnetic field \textbf{B} can be described by the following term included in the Dirac-Coulomb Hamiltonian:
\begin{equation}
 \label{HB}
{\rm H}_B={\rm \bf{B}}\cdot \frac{|e|}{2}(\bm{r}_G \times \bm{\alpha}),
\end{equation}
where $\bm{\alpha}$ are the Dirac matrices and $\bm{r}_G = \bm{r} - \bm{R}_G$, $\bm{R}_G$ is the gauge origin~\cite{DyallFaegri2007}, i.e. the origin for the coordinate system that describes the electron radius-vector in this equation. In principle, the choice of $\bm{R}_G$ can influence the results obtained in the modest basis sets (see below). 
In the point magnetic dipole approximation, the hyperfine interaction of an electron with the magnetic moment $\bm{\mu}_j$ of the $j$th nucleus can be written in the following way:
\begin{equation}
 \label{HHFS}
{\rm H}_{\rm hyp}=\frac{|e|}{4\pi} \bm{\mu}_j\cdot \frac{(\bm{r}_j \times \bm{\alpha})}{r_j^3},
\end{equation}
where $\bm{r}_j=\bm{r} - \bm{R}_j$, $\bm{R}_j$ is the position of the nucleus $j$.
Note, that the interaction (\ref{HHFS}) does not take into account the finite nuclear magnetization distribution effect. In the theory of atomic hyperfine structure this effect is called the Bohr-Weisskopf (BW) effect~\cite{bohr1950influence, bohr1951bohr, sliv1951uchet}. One can use the following substitution to consider this effect~\cite{Zherebtsov:2000,shabaev1997ground,Tupitsyn:02}:
\begin{equation}
\label{muF}
\bm{\mu}\to\bm{\mu}(r)=\bm{\mu} F(r).
\end{equation}
Function $F(r)$ takes into account the nuclear magnetization distribution inside the finite nucleus. In the point magnetic dipole moment approximation $F(r)=1$. In the finite distribution case $F(r)$ can significantly differ from $1$ inside the nucleus. Expressions for different models can be found in Refs.~\cite{Zherebtsov:2000,Tupitsyn:02,Volotka:2008,Malkin:2011,Roberts:2021}. In the simplest uniformly magnetized ball model function $F(r)=(r/r_n)^3$ for $r$ inside the sphere of radius $r_n=\sqrt{5/3}r_c$ ($r_c$ is the root-mean-square charge radius) and is equal to $1$ outside~\cite{Volotka:2008}.
In studies of neutral atoms, this model is most widely used to calculate the BW correction~\cite{Sapirstein:2003,konovalova2017calculation,Prosnyak:2020, Ginges:2018,kozlov2001parity}. In the present paper, we mainly use the model which implies that magnetization can be ascribed to the single-particle structure of the nucleus. 
In this model function $F(r)$ is given by~\cite{Zherebtsov:2000}:
\begin{equation}
\begin{split}
F(r') = \frac{\mu_N}{\mu}\left\{ \int_0^{r'}r^2|u(r)|^2dr\left[\frac{1}{2}g_S+
\right. \right. \\ \left. \left. 
\left(I-\frac{1}{2}+ 
\frac{2I+1}{4(I+1)}m_p\phi_{SO}(r)r^2\right)g_L\right] + 
\right. \\ \left. 
\int_{r'}^{\infty} r^2\left(\frac{r'}{r}\right)^3|u(r)|^2dr\left[-\frac{2I-1}{8(I+1)}g_S + 
\right. \right. \\ \left. \left.
\left(I-\frac{1}{2} +\frac{2I+1}{4(I+1)}m_p\phi_{SO}(r)r^2\right)g_L
\right]\right\}
\end{split}
\end{equation}
for $I=L+1/2$, and 
\begin{equation}
\begin{split}
F(r') = \frac{\mu_N}{\mu}\left\{\int_0^{r'} drr^2|u(r)|^2 \left[-\frac{I}{2(I+1)}g_S+ \right. \right. \\ \left. \left.
\left(\frac{I(2I+3)}{2(I+1)}-\frac{2I+1}{4(I+1)}m_p\phi_{SO}(r)r^2\right)g_L\right] + \right.\\
\left. \int_{r'}^{\infty} r^2\left(\frac{r'}{r}\right)^3|u(r)|^2dr \left[\frac{2I+3}{8(I+1)}g_S+
\right. \right. \\ \left. \left.
\left(\frac{I(2I+3)}{2(I+1)}- 
\frac{2I+1}{4(I+1)}m_p\phi_{SO}(r)r^2\right)g_L\right]\right\}
\end{split}
\end{equation}
for $I=L-1/2$. Here $\mu_N$ is the nuclear magneton, $m_p$ is the proton mass, $I$ is the nuclear spin, $\left|u(r)\right|^{2}$ is the density of the valence nucleon, $\phi_{SO}$ is the radial part of the spin–orbit interaction $V_{SO} = \phi_{SO} \: \pmb{ \sigma }\cdot \pmb{l}$, $\pmb{l}$ is the angular moment operator and $\pmb{ \sigma }$ is the vector of Pauli matrices. In the Woods-Saxon (WS) model of the nucleus, the wave function of the valence nucleon is determined as a solution of the Schr{\"o}dinger equation with the WS potential. A detailed description of the implementation
and parameters of the potential can be found in Ref.~\cite{Prosnyak:2021} and references therein. For the valence proton we set $g_L=1$, for the valence neutron $g_L=0$. 
Parameter $g_S$ is obtained from the following equations:
\begin{equation}
    \frac{\mu}{\mu_N} = \frac{1}{2}g_S + \bigg[I-\frac{1}{2}+\frac{2I+1}{4(I+1)}m_p\langle \phi_{SO}r^2\rangle\bigg]g_L
\label{gsLm}         
\end{equation}
for $I=L+1/2$, and
\begin{equation}
    \frac{\mu}{\mu_N} = -\frac{I}{2(I+1)}g_S + \bigg[\frac{I(2I+3)}{2(I+1)}-\frac{2I+1}{4(I+1)}m_p\langle \phi_{SO}r^2\rangle\bigg]g_L
\label{gsLp}      
\end{equation}
for $I=L-1/2$. In the simple single-particle model with the uniform distribution of the valence nucleon, the density of the valence nucleon $|u(r)|^2$ is a constant inside the nucleus volume and there is no spin-orbit term in this model~\cite{shabaev1994hyperfine}.

In the one-electron case, the tensor (\ref{SHIELDINGDer}) can be calculated using the sum-over-states method corresponding to the second-order perturbation theory with perturbations (\ref{HB}) and (\ref{HHFS}):
 \begin{eqnarray}
 \label{SHIELDPT}
& & \sigma_{a,b}=\\ 
& & \sum_{n \neq 0}
 \frac{\langle 0 | \frac{|e|}{4\pi} (\frac{(\bm{r}_j \times \bm{\alpha})}{r_j^3})_a  |n \rangle
       \langle n|   (\frac{|e|}{2}(\bm{r}_G \times \bm{\alpha}))_b     |0 \rangle}
      {E_0-E_n} + h.c. \nonumber,
 \end{eqnarray}
where $|0 \rangle$ is the unperturbed one-particle state of interest, $|n \rangle$ is the unoccupied n'th unperturbed state (orbital) and h.c. is the Hermitian conjugate. 
For the case of the four-component Dirac theory, the summation should include both positive energy and negative energy states $|n \rangle$~\cite{Aucar:99}. The part of the sum associated with the positive energy states is called ``paramagnetic'' term. The part associated with the negative energy states is called ``diamagnetic term''~\cite{Aucar:99}. For the cases of the Dirac-Hartree-Fock (DHF) and density functional theory (DFT) many-electron methods, one can use the response technique to calculate both terms~\cite{Olejniczak:12,Ilias:13,Aucar:99,DIRAC15}. The result of the application of this technique is equivalent to (and derived from) the analytical calculation of the DHF/DFT energy derivative (\ref{SHIELDINGDer}). In the present paper we have used the implementation of the method within the {\sc dirac} \cite{DIRAC15,Saue:2020} code.

In calculations of the shielding constant, we have used the following Gaussian-type basis sets to describe electronic wave functions. The first one corresponds to the uncontracted Dyall's
AE4Z~\cite{Dyall:07,Dyall:12} basis set for all atoms and will be called QZQZ below. 
This basis set contains $[34s\, 30p\, 19d\, 14f\, 10g\, 5h\, 1i]$ primitive Gaussian functions for Re and $[18s\, 10p\, 5d\, 3f\, 1g]$ functions for each oxygen.
The second one, TZTZ, corresponds to the uncontracted AE3Z~\cite{Dyall:07,Dyall:12} basis set on rhenium and contracted aug-cc-pVTZ~\cite{Dunning:89,Kendall:92} on oxygen. 
This basis set contains  $[30s\, 24p\, 15d\, 11f\, 5g\, 1h]$ functions for Re and $[5s\, 4p\, 3d\, 2f]/([11s\, 6p\, 3d\, 2f)$  for each oxygen, where in the $[...]$-brackets the numbers of contracted functions are given and in the $(...)$-brackets the corresponding numbers of primitive functions are given (e.g. each of five contracted $s-$type functions of oxygen is a linear combination of 11 primitive functions). We have also used the DZDZ basis set which corresponds to the uncontracted Dyall's AE2Z~\cite{Dyall:07,Dyall:12} basis set on rhenium and aug-cc-pVDZ~\cite{Dunning:89,Kendall:92} on oxygen.
This basis set contains  $[24s\, 19p\, 12d\, 9f\, 1g]$ functions for Re and $[4s\, 3p\, 2d]/([10s\, 5p\, 2d)$ for oxygen. The quality of basis sets increases in the series: DZDZ, TZTZ, QZQZ.

Formally, the interaction of a molecule with an external uniform magnetic field should not depend on the choice of the origin $\bm{R}_G$ in Eq.~(\ref{HB}). But for finite-size basis sets there may be some dependence~\cite{DyallFaegri2007,Olejniczak:12,Ilias:13} which can affect the shielding constant value. To minimize such a dependence one can use the London atomic orbitals (LAOs) method, developed at the four-component DFT level in Refs.~\cite{Olejniczak:12,Ilias:13}. In this approach, basis functions are replaced by the so-called London atomic orbitals which are obtained from the original basis functions by applying a magnetic field-dependent factor~\cite{Olejniczak:12,Ilias:13,DyallFaegri2007}. This corresponds to the transformation of the wave function due to the gauge transformation of the vector potential in Eq.~(\ref{HB}). The use of London orbitals guarantees the gauge-origin invariance of results in a finite basis approximation~\cite{Olejniczak:12,Ilias:13,DyallFaegri2007}. Even for usual basis set, the gauge-origin  problem should decrease with the basis set size increase. In the present case, we are interested in the shielding constant for the rhenium nucleus. Therefore, it is natural to place the origin at this nucleus. According to our DFT estimates, the values of the shielding constant calculated for the QZQZ basis set (i) with such choice of the origin and employing usual basis functions or (ii) within the LAOs technique coincide within 7~ppm. This value is negligible in comparison with the total uncertainty of the present calculation (see below).

Geometry structure parameters of the ReO$_4^-$ anion have been optimized using the four-component density functional theory with the Perdew-Burke-Ernzerhof, PBE0, functional~\cite{pbe0} and using the TZTZ basis set. No solvent effects were considered at this stage. The optimized value of the Re--O bond length in the ReO$_4^-$ anion with the regular tetrahedral symmetry was found to be 1.723\AA. This value is in good agreement (within 0.003\AA) with the  study~\cite{Antusek:20}. 

Relativistic four-component calculations were performed within the locally modified {\sc dirac15} \cite{DIRAC15,Saue:2020} code. High-order correlation effects have been calculated using the {\sc mrcc} code \cite{MRCC2020}. The code for calculating the BW matrix elements in the WS model has been developed in Ref.~\cite{Prosnyak:2021} for atoms and generalized to the molecular case in the present paper. Taking into account that the action of the corresponding operator is localized inside the nucleus we have neglected the contribution of basis functions centered on oxygen atoms in the present implementation. In the molecular electronic structure calculations, the Gaussian nuclear charge distribution model~\cite{Visscher:1997} has been used. 

\section{Results and discussion}

\begin{table}[]
\caption{Calculated values of rhenium shielding constant $\sigma$ contributions for ReO$_4^{-}$ in ppm.}
\label{ShieldCalc}
\begin{tabular}{lr}
\hline
\hline
Contribution &  Value \\
\hline

Diamagnetic:  \\
~~QZQZ-LAO/PBE0       & 7633  \\
Paramagnetic: \\
~~TZTZ/108e-CCSD         & $-3741$  \\
~~TZTZ/108e-CCSD(T) - 108e-CCSD~~ & 350   \\
~~DZDZ/24e-CCSDT - 24e-CCSD(T)~~ & -81   \\
~~Basis set correction   & $-10$   \\
Gaunt                          & 15   \\
Solvent effect, from Ref.~\cite{Antusek:20} & $-25$   \\
Finite magn. distribution (WS)    &  $-73$  \\
\\
Total                  & 4069 \\
\hline
\hline
\end{tabular}
\end{table}

\subsection{\label{sectShieldCalc} Shielding constant calculation}
We have used the following scheme to calculate the shielding constant for ReO$_4^-$ and its contributions (see Table~\ref{ShieldCalc}). The diamagnetic part has been calculated at the four-component PBE0 method~\cite{pbe0}. As in previous studies~\cite{Skripnikov:18a,Skripnikov:2020a} we have found that this contribution is almost independent of the choice of the functional or method used. For example, the values calculated within the Dirac-Hartree-Fock (7630.8 ppm) and PBE0 (7633.3 ppm) response theories coincide within a few ppm. Moreover, the same value within a few ppm can be obtained even using the uncoupled Dirac-Hartree-Fock approach (7633.2 ppm), i.e. simple orbital perturbation theory. The latter corresponds to calculation using Eq.~(\ref{SHIELDPT}) with an additional summation over all occupied molecular orbitals, $|0 \rangle$, which are included in the Slater determinant, and the sum over $n$ in Eq.~(\ref{SHIELDPT}) is limited to the negative energy orbitals.

The most challenging part of the problem is the calculation of the paramagnetic contribution to the shielding constant which is strongly affected by both correlation and relativistic effects. The authors of Ref.~\cite{Antusek:20} have used the \textit{nonrelativistic} coupled cluster theory combined with the relativistic correction calculated within the DFT approach to calculate the shielding constant for ReO$_4^-$. As it has been analysed in Ref.~\cite{Antusek:20} the dominant source of the uncertainty of such an approach is ``the systematic error of correlation and relativistic effects nonadditivity''. 
To avoid such an error, we avoided the use of non-relativistic theory at all and used the four-component \textit{relativistic} coupled clusters theory as the main approach for calculating the paramagnetic contribution to the shielding constant.
We have directly calculated the mixed derivative  (\ref{SHIELDINGDer}) within the standard numerical finite-difference technique. For this we needed the (numerical) dependence of the energy $E$ on the magnetic moment and external magnetic field values. To calculate this dependence we have added perturbations (\ref{HB}) and (\ref{HHFS}) with required (small) values of the nuclear magnetic moment and the external uniform magnetic field amplitude to the molecular relativistic Hamiltonian and solved coupled cluster equations with these perturbed Hamiltonian to obtain perturbed values of the energy $E$. 
This electronic correlation calculation has been performed within the TZTZ basis set using the relativistic coupled cluster with single, double and perturbative triple cluster amplitudes method, CCSD(T)\cite{Visscher:96a}. All 108 electrons of ReO$_4^-$ were included in the correlation calculation and no virtual energy cutoff has been applied. In Table~\ref{ShieldCalc} we separate the CCSD(T) value into the CCSD value (line ``TZTZ/108e-CCSD'') and a contribution of perturbative triple cluster amplitudes (line ``TZTZ/108e-CCSD(T) - 108e-CCSD''). Thus the ``TZTZ/108e-CCSD(T) - 108e-CCSD'' line gives the difference of the shielding constants calculated within the CCSD(T) and CCSD methods (in both cases all electrons were correlated and the TZTZ basis set has been used).

To explore even higher-order correlation effects, we have performed correlation calculations within the relativistic coupled cluster with single, double and \textit{iterative} triple cluster amplitudes method, CCSDT, and compared it with the CCSD(T) one. Due to extremely high complexity of the CCSDT approach (e.g. in the present calculation our cluster operator included 2.3 billions cluster amplitudes), we have included 24 valence electrons of ReO$_4^-$ in these two calculations and employed the DZDZ basis set. One should note that the contribution of perturbative triple cluster amplitudes in the DZDZ/24e-CCSD(T) calculation reproduces such contribution obtained in the main calculation (350 ppm, see line ``TZTZ/108e-CCSD(T) - 108e-CCSD'' in Table~\ref{ShieldCalc}) within 80\%. As one can see, the difference between CCSDT and CCSD(T) results (-81 ppm) is rather small (see the ``DZDZ/24e-CCSDT - 24e-CCSD(T)'' line in Table \ref{ShieldCalc}).

To take into account the effect of the extended basis set with respect to the main TZTZ one, we have calculated basis set correction within the relativistic PBE0 approach. In this calculation, we have increased the basis set up to the QZQZ one, i.e. we have calculated the difference of the paramagnetic contributions to shielding constants calculated within the QZQZ and TZTZ basis set using the PBE0 approach. One can see from Table~\ref{ShieldCalc} that this correction is small (-10 ppm). Note, that there is no guarantee, that DFT can reasonably take into account basis set correction accurately. Therefore, we have estimated the influence of the basis set size increase from the DZDZ basis set to the TZTZ one on the paramagnetic part of the shielding constant. Here we studied how DFT (PBE0)  can reproduce this effect, calculated within the wave function-based relativistic CCSD(T) method. Obtained corrections are: -30 ppm within DFT vs. -71 ppm within CCSD(T). As expected, DFT underestimated the effect of the basis set size increase (by a factor of 2.4). Thus, the mentioned correction (-10 ppm) in Table~\ref{ShieldCalc} can be underestimated. We take this fact into account in the uncertainty estimation below.

The effect of solvent has been extensively analyzed in Ref.~\cite{Antusek:20}. It seems that the ``$\delta_4$'' scheme used in Ref.~\cite{Antusek:20} is the most elaborate study of this effect at present. It explicitly takes into account the effect of the first solvation shell and approximately takes into account the influence of the solution on the shielding constant under consideration (within the polarizable continuum model) at the DFT level~\cite{Antusek:20}. Therefore, we include this contribution in our final value. 

The Gaunt interaction contribution has been calculated at the relativistic DFT (PBE0) level using the TZTZ basis set, i.e. it has been calculated as a difference between the shielding constant values obtained within the relativistic DFT method with inclusion and without inclusion of the Gaunt interaction into the electronic Hamiltonian.

In the present paper, we have studied the influence of the finite nuclear magnetization distribution on the shielding constant. For this, we have used the substitution given by Eq.~(\ref{muF}) in the hyperfine interaction operator~(\ref{HHFS}) and have used the single-particle WS model, described above. Calculation of the paramagnetic part has been performed at the relativistic CCSD(T) level using the DZDZ basis set, while the diamagnetic part has been calculated at the uncoupled Dirac-Hartree-Fock level (this method is described above). The latter contribution, 8 ppm, to the considered correction was found to be much smaller than the paramagnetic part, $-81$ ppm, of this correction. As one can see from Table~\ref{ShieldCalc} the finite nuclear magnetization distribution effect (-73 ppm or 1.8\% of the total $\sigma$) is more important than the solvent effect for the system under consideration. For both isotopes considered, nuclear magnetization distribution effects were found to be almost identical and they are not distinguished in Table~\ref{ShieldCalc}. We did not find previous attempts to take into account the influence of the finite nuclear magnetization distribution effect on the shielding constants in many-electron molecules, although such studies have been performed for H-like ions~\cite{Yerokhin:2011,Yerokhin:2012}.

\subsection{\label{sectUncert}Shielding constant uncertainty estimation}
The total uncertainty ($\delta$) of the present calculation of shielding constant can be estimated as a square root of the sum of squares of electronic correlation calculation uncertainty ($\delta_{\rm el. corr.}$), basis set uncertainty ($\delta_{\rm BS}$), Gaunt interaction effects inclusion uncertainty ($\delta_{\rm G}$), finite nuclear magnetization distribution uncertainty ($\delta_{\rm FMD}$), quantum electrodynamics ($\delta_{\rm QED}$), uncertainty due to geometry structure uncertainty ($\delta_{\rm geom}$) and solvent effects uncertainty ($\delta_{\rm sol}$):
\begin{equation}
\label{uncerts}
    \delta=\sqrt{\delta_{\rm el. corr.}^2 + \delta_{\rm FMD}^2 + \delta_{\rm G}^2 + \delta_{\rm BS}^2 + \delta_{\rm sol}^2 + \delta_{\rm QED}^2 + \delta_{\rm geom}^2 }.
\end{equation}

Uncertainty of the calculated electronic correlation effect $\delta_{\rm el. corr.}$ can be estimated as the contribution of the perturbative triple cluster amplitudes given in Table \ref{ShieldCalc} in the line ``TZTZ/108e-CCSD(T) - 108e-CCSD'', i.e. $\delta_{\rm el. corr.}=$350 ppm. For a more detailed analysis of this contribution, we also calculated it within the smaller DZDZ basis set, i.e. calculated the difference between shielding constants calculated using the DZDZ/108e-CCSD(T) and DZDZ/108e-CCSD approaches. The calculated value, 355~ppm, reproduces with high accuracy the contribution of perturbative triple clusters amplitudes within the TZTZ basis set given above, 350~ppm. This suggests a small contribution from the interference effect between the high-order correlation effects defined by perturbative triple amplitudes and the size of the basis set (the uncertainty associated with the size of the basis set itself is discussed below). Note that the indicated uncertainty $\delta_{\rm el. corr.}$ seems to be quite conservative, given that the estimate of the contribution of higher-order correlation effects (outside the considered CCSD(T) approximation) calculated within the framework of the CCSDT approach (see ``DZDZ/24e-CCSDT - 24e-CCSD(T)'' line in Table \ref{ShieldCalc} and description in section~\ref{ShieldCalc}) is several times smaller than $\delta_{\rm el. corr.}$.

The uncertainty $\delta_{\rm FMD}$ of the calculated value of the finite nuclear magnetization distribution contribution (-73 ppm) can be estimated as about 30\% of the value of this contribution (see also the analysis of the HFS data for H-like Re below): $\delta_{\rm FMD}=23$~ppm. This value is obtained by comparing contributions of the finite nuclear magnetization distribution to the shielding constant calculated in the WS model, -73~ppm, and in the uniformly magnetized ball model, -96~ppm. The latter value has been calculated using the same approach as that employed for the first one (described in the previous section) with only replacement of the nuclear magnetization distribution function $F(r)$.

As a measure of the uncertainties of the Gaunt interaction and solvent effects calculated within one method (DFT), we have used corresponding values of these effects given in Table \ref{ShieldCalc}, i.e. $\delta_{\rm G}=15$~ppm, $\delta_{\rm sol}=25$~ppm. This means that the (conservative) uncertainties of these corrections are suggested to be 100\%. A more accurate calculation of these effects and their uncertainties will be required when the remaining uncertainties are made smaller (e.g. the uncertainty $\delta_{\rm el. corr.}$ considered above is more than an order of magnitude bigger than the effects under consideration).

Let us estimate the uncertainty due to the basis set  incompleteness $\delta_{\rm BS}$. For this we can compare shielding constant calculated within the best employed QZQZ basis set and a smaller one, TZTZ basis set. 
These differences for the diamagnetic contributions, paramagnetic contribution and the total shielding constant values calculated within the QZQZ and TZTZ basis sets using the DFT (PBE0) method are $+$38~ppm, $-$10~ppm and $+ $28~ppm, respectively. Thus, we can suggest that $\delta_{\rm BS}$ is about 28~ppm. However, as we have mentioned above, the DFT approach can underestimate the effect of the basis set size increase for the paramagnetic contribution by a factor of 2.4. This will lead to the estimation for the uncertainty of the basis set incompleteness as $38-10\cdot2.4=$14~ppm. 
For the conservative estimate we choose the largest of these two possible values, i.e. we set $\delta_{\rm BS}=28$~ppm.

The geometry parameters, i.e. the Re--O distances in the ReO$_4^-$ anion have been optimized at the relativistic PBE0 level (see above). To check the uncertainty of the optimized geometry we have also performed geometry optimization within another popular functional -- B3LYP~\cite{b3lyp}. Within this approach the optimized Re--O distance was found to be 1.736~\AA, i.e. the estimation for the geometry parameters uncertainty of ReO$_4^-$ is about 0.013~\AA. According to our DFT-based estimation, this uncertainty in the geometry structure parameters leads to contribution to the uncertainty of the shielding constant of about $\delta_{\rm geom}=158$~ppm. This is about six times bigger than the solvent effect contribution (and is about an order of magnitude bigger than the Gaunt interaction effect).

In a recent paper~\cite{Koziol:2019} it was estimated that contribution of quantum electrodynamics effects to the shielding constant is about 0.5\% for such many-electon atoms as astatine (atomic number 85). 
For H-like ions, ab-initio calculations are available~\cite{Yerokhin:2011,Yerokhin:2012}. According to these Refs., the QED contribution to the shielding constant of H-like Bi (atomic number 83) is 0.7\% and generally increases with atomic number. In the present work, we do not take into account QED effects but according to the notes above suggest that their contribution can be about 1\% of the total value of the shielding constant and include this value in the uncertainty, i.e $\delta_{\rm QED}$=41 ppm.

The final value of the uncertainty of the shielding constant is dominated by the correlation contribution uncertainty $\delta_{\rm el. corr.}$. Substituting all estimated values of the uncertainties in Eq.~(\ref{uncerts}) we obtain the total uncertainty value: $\delta$=389~ppm, i.e. the final value of the shielding constant is $\sigma=$4069(389)~ppm. It is in reasonable agreement with the previous calculation 3698(927)~ppm~\cite{Antusek:20} but has reduced uncertainty due to the use of the relativistic coupled cluster approach for the most challenging part of the calculation. Note also, that in the present paper we consider more sources of the uncertainties.

\subsection{\label{sectNMRmoments}New values of magnetic dipole moments of $^{185}$Re and $^{187}$Re}
To obtain the nuclear magnetic dipole moments of $^{185}$Re and $^{187}$Re we need corresponding uncorrected values of these moments, i.e. magnetic dipole moments which are not corrected for the magnetic shielding, see Eq.~(\ref{mumuuncorr}). The nuclear magnetic resonance experiment on the ReO$_4^-$ anion has been performed in the aqueous solution of NaReO$_4$ with the magnetic dipole moment of $^{23}$Na as the reference~\cite{Alder:1951}. In the experiment the two resonances of $^{185}$Re and $^{187}$Re were located near a frequency of 6.4 Mc in an external field of 6700 gauss. The following results for the resonance frequencies $\nu$ were obtained in the experiment~\cite{Alder:1951}:
\begin{eqnarray}
  \nu(^{185}{\rm Re})/\nu(^{23}{\rm Na})=0.85114(9), \\
  \nu(^{187}{\rm Re})/\nu(^{23}{\rm Na})=0.85987(9). 
\end{eqnarray}
As in Ref.~\cite{Gustavsson:1998} we use the uncorrected NMR value $\mu^{\rm uncorr.}(^{23}{\rm Na})$ from Ref.~\cite{Lutz:1967} to obtain the uncorrected values of the nuclear magnetic moments of $^{185}$Re and $^{187}$Re~\cite{Gustavsson:1998} according to the relation~\cite{Gustavsson:1998} $\mu^{\rm uncorr.}({\rm Re})=\mu^{\rm uncorr.}({\rm Na}) (\nu({\rm Re})/\nu({\rm Na}))(I_{\rm Re}/I_{\rm Na})$ (where the ${\rm ^{23}Na}$ nuclear spin $I_{\rm ^{23}Na}$=1.5):
\begin{eqnarray}
     \mu^{\rm uncorr.}(^{185}{\rm Re})=3.1439(3) \mu_N, \\
     \mu^{\rm uncorr.}(^{187}{\rm Re})=3.1761(3) \mu_N.  
\end{eqnarray}
Using these uncorrected values and our theoretical shielding constant value $\sigma$ we obtain the \textit{final values} of the nuclear magnetic moments according to Eq.~(\ref{mumuuncorr}):
\begin{eqnarray}
\label{NMRvals185}
     \mu(^{185}{\rm Re})=3.1567(3)(12)~\mu_N, \\
\label{NMRvals187}     
     \mu(^{187}{\rm Re})=3.1891(3)(12)~\mu_N.  
\end{eqnarray}
Here the first uncertainty is due to the experiment and the second is due to the present theory.

The obtained value of the shielding constant for ReO$_4^-$ molecular anion is about three times smaller than the shielding constant for the Re$^{7+}$ atomic  cation~\cite{Johnson:68,Feiock:1969,Kolb:1982} which was used in some of the previous interpretations of the molecular NMR data~\cite{stone2014,Raghavan:89,Gustavsson:1998}. It means that in previous studies the uncertainty of the shielding correction used for interpretation of the molecular NMR experiment has been substantially underestimated~\cite{Gustavsson:1998}.

\section{\label{sectHlike}Hydrogen-like rhenium ions}
\subsection{Nuclear magnetic moments from HFS data for H-like Re}
Rhenium is one of several elements for which measurements of hyperfine splitting for H-like (Re$^{74+}$) ion was carried out~\cite{Crespo:1998}. In principle, it is possible to extract the magnetic moment value from these HFS data if the values of BW and QED contributions are known~\cite{CrespoBeiersdorfer97}. For this one can use the following expression for the hyperfine structure constant $A$
~\cite{Note1}:
\begin{equation}
    A = A^{(0)} - A^{\rm BW}+A^{\rm QED} = A^{(0)}(1-\varepsilon)+A^{\rm QED},
\label{AA0}    
\end{equation}
where $A^{(0)}$ is the HFS constant calculated in the point magnetic dipole approximation, $A^{\rm BW}$ is the Bohr-Weisskopf contribution to the HFS constant, $\varepsilon$ is the relative Bohr-Weisskopf correction, and $A^{\rm QED}$ is the QED contribution. Constants $A^{(0)}, A^{\rm BW}$ and $A^{\rm QED}$ are proportional to the nuclear magnetic moment $\mu=g_I I \mu_N$, where $g_I$ is the g-factor of nucleus with spin $I$. $A^{(0)}$ and $A^{\rm QED}$ can be accurately calculated~\cite{shabaev1997ground,artemyev2001vacuum} and $A^{\rm BW}$ can be estimated within some nuclear magnetization distribution model.
In Ref.~\cite{Prosnyak:2021} we have calculated the relative BW correction $\varepsilon$ for H-like $^{185}$Re. Here we have also estimated BW effect for the $^{187}{\rm Re}$ isotope using four different nuclear magnetization distribution models (see Table~\ref{Eps}): uniformly magnetized ball model (Ball), single-particle model with the uniform distribution of the valence nucleon (UD), WS model with and without spin-orbit interaction in Schr{\"o}dinger equation for valence nucleon (see the Theory section). The obtained values for both considered isotopes are almost identical, since these nuclei have similar single-particle structures and charge radii. Using the experimental values of $A$~\cite{Crespo:1998}, the values of BW corrections $\varepsilon$ calculated within different nuclear magnetization distribution models and given in Table~\ref{Eps}, the values of the ratio $A^{(0)}/g_II=0.2926(3)$~eV calculated in~\cite{shabaev1997ground,shabaev1994hyperfine}
~
\cite{Note2}
and QED effect calculated in Refs.~\cite{shabaev1997ground,artemyev2001vacuum} 
$A^{\rm QED}/g_II=-0.00158(3)$~eV
we have determined the corresponding values of magnetic moments according to the equation $\mu=A/[A^{(0)}(1-\varepsilon)/g_II+A^{\rm QED}/g_II]$. The values of $\mu$ deduced in such a way using different models of nuclear magnetization distribution are given in Table~\ref{HFS_mu}. In such approach the main uncertainty is due to the BW effect, as it is hard to reliably treat many-body nuclear structure effects. We suppose that this uncertainty can be estimated by comparing different nuclear magnetization distribution models given in Table~\ref{Eps}. 
\begin{table}[]
\caption{Calculated values of the relative BW correction $\epsilon$ to hyperfine structure constants of H-like rhenium in different nuclear models in \%. Both rhenium isotopes $^{185}{\rm Re}$ and $^{187}{\rm Re}$ have nuclear spin $I=2.5$.}
\label{Eps}
\begin{tabular}{lrr}
\hline
\hline
 &  $^{185}{\rm Re}$ & $^{187}{\rm Re}$ \\
\hline
Ball & 1.69 & 1.69 \\
UD & 1.35 & 1.36 \\
WS without SO & 1.30 & 1.30 \\
WS with SO & 1.32 & 1.32 \\
\hline
\hline
\end{tabular}
\end{table}
Using this approach, the relative Bohr-Weisskopf correction can be estimated as $\varepsilon=$1.32(37)\% for both isotopes
~\cite{Note3}.
It corresponds to the following value of the magnetic moments derived from the experimental HFS data~\cite{Crespo:1998} for H-like ions:
\begin{eqnarray}
\label{muHFS185}
     \mu^{\rm (HFS)}(^{185}{\rm Re})= 3.156(2)(3)(12)~\mu_N, \\
\label{muHFS187}     
     \mu^{\rm (HFS)}(^{187}{\rm Re})= 3.187(2)(3)(12)~\mu_N.       
\end{eqnarray}
Here the first uncertainty is due to the experimental determination of $A$~\cite{Crespo:1998}, the second one is due to uncertainties of $A^{(0)}$
(the uncertainty of $A^{\rm QED}$ is negligible) and the third one corresponds to the uncertainty of the calculated BW effect. These values are in agreement with Ref.~\cite{CrespoBeiersdorfer97} and in agreement with the values (\ref{NMRvals185}) and (\ref{NMRvals187}) derived from the NMR data above but have an order of magnitude larger uncertainty. It is mainly due to the BW effect uncertainty. The influence of the BW effect on the shielding constant (1.8\%) in the considered molecular anion and on the hyperfine structure of H-like rhenium ion (1.3\%) are similar (see also Tables \ref{ShieldCalc} and \ref{Eps}). However, the influence of the BW effect on the final value of the magnetic moment, extracted from the NMR data is much smaller than in the case of H-like HFS data. In the first case, the finite nuclear magnetization distribution effect is a correction to the shielding effect, which is 4069 ppm, i.e. only 0.4\% itself (see above). The uncertainty of the molecular electronic structure shielding constant calculation can be controlled, see section \ref{sectUncert}. In the second case, the BW effect directly contributes to the magnetic moment, and an estimation of its uncertainty is complicated due to the absence of direct many-body calculations of this effect for the rhenium nucleus. 

It is possible to employ the obtained results also as follows. If one uses the tabulated values~\cite{stone2014} of the nuclear magnetic moments of $^{185}$Re and $^{187}$Re, the following theoretical values of the HFS constant $A^{\rm th}$ for H-like rhenium can be obtained according to the equation $A=\mu[A^{(0)}(1-\varepsilon)/g_II+A^{\rm QED}/g_II]$: $A^{\rm th}(^{185}{\rm Re})=0.9152(34)$~eV and $A^{\rm th}(^{187}{\rm Re})=0.9245(34)$~eV. 
For the new values of magnetic moments (\ref{NMRvals185}), (\ref{NMRvals187}), extracted from the NMR data above one obtains:  $A^{\rm th}(^{185}{\rm Re})=0.9065(34)$~eV and $A^{\rm th}(^{187}{\rm Re})=0.9157(34)$~eV. The experimental values $A^{\rm exp}$ are~\cite{Crespo:1998}: $A^{\rm exp}(^{185}{\rm Re})=0.9063(6)$ eV and $A^{\rm exp}(^{187}{\rm Re})=0.9150(6)$ eV. Thus, the updated values of the nuclear magnetic moments resolve the disagreement between theoretical predictions of HFS constants (or HFS splittings) of H-like rhenium ions (based on the old values of magnetic moments)~\cite{shabaev1997ground,beier2000gj,boucard2000relativistic} and experimental values~\cite{Crespo:1998}.

\begin{table}[]
\caption{Values of the nuclear magnetic moments (in units of $\mu_N$) extracted from the experimental data on hyperfine structure constants of H-like rhenium~\cite{Crespo:1998} using QED corrections from Refs.~\cite{shabaev1997ground,artemyev2001vacuum} and calculated BW corrections within different nuclear magnetization distribution models.
For an uncertainty estimation, see the main text.}
\label{HFS_mu}
\begin{tabular}{lrr}
\hline
\hline
 &  $^{185}{\rm Re}$ & ~~~~~~$^{187}{\rm Re}$ \\
\hline
 Ball          & 3.168 & 3.199 \\
 UD            & 3.157 & 3.188 \\
 WS without SO & 3.155 & 3.186 \\
 WS with SO    & 3.156 & 3.187 \\
\hline
\hline
\end{tabular}
\end{table}

\subsection{BW effect from HFS data for H-like ions and magnetic moments from molecular NMR data}
As mentioned in the Introduction, accurate theoretical prediction of the hyperfine structure of atoms and molecules can be used to probe the accuracy of the electronic wave function. However, such predictions depend on the nuclear magnetic dipole moment value and the function of the nuclear magnetization distribution $F(r)$ in Eq.~(\ref{muF}). If both of these components are accurately known, then one can predict the hyperfine structure of the compound or ion under consideration. It was shown in Ref.~\cite{Skripnikov:2020e} that for many-electron heavy-atom molecules and heavy atoms to a good approximation it is possible to factorize the Bohr-Weisskopf contribution to the hyperfine structure constant into a pure electronic part and just one universal numerical parameter, which depends on the nuclear magnetization distribution, see Eq.~(29) in Ref.~\cite{Skripnikov:2020e} (recently a related approach has been considered for atoms in $s$ and $p_{1/2}$ states in Ref.~\cite{Ginges:2022}). The latter parameter $B_s$~\cite{Skripnikov:2020e} is proportional to the BW contribution  $A^{\rm BW}$ to the hyperfine structure constant in Eq.~(\ref{AA0}) for the H-like ion in the ground electronic state and can be calculated as $B_s=A^{\rm BW}/2g_I$ in this case; actually, the constant of interest is the product $B_Sg_I=A^{\rm BW}/2$~\cite{Skripnikov:2020e}. Using the \textit{experimental} values of the H-like rhenium HFS constants~\cite{Crespo:1998}, QED corrections from Refs.~\cite{shabaev1997ground,artemyev2001vacuum} and the values of the nuclear magnetic moments (\ref{NMRvals185}) and (\ref{NMRvals187}) refined in the present paper above, we obtain the following values of the BW contribution to the HFS constants, $A^{\rm BW (``exp'')}$, according to Eq.~(\ref{AA0}):
\begin{eqnarray}
\label{ABW185}
     A^{\rm BW (``exp'')}(^{185}{\rm Re})=0.0124(6)(8)(4) {\rm~eV}, \\
\label{ABW187}     
     A^{\rm BW (``exp'')}(^{187}{\rm Re})=0.0130(6)(9)(4) {\rm~eV}.
\end{eqnarray}
Here the first uncertainty is due to the experimental HFS data for H-like ions~\cite{Crespo:1998}, the second one is due to uncertainty of $A^{(0)}$ and the third one corresponds to nuclear magnetic moment uncertainties in (\ref{NMRvals185}) and (\ref{NMRvals187}). From these values one can obtain for the product $B_Sg_I=A^{\rm BW}/2$: 
$B_Sg_I(^{185}{\rm Re})=0.0062(3)(4)(2) {\rm~eV}$,
$B_Sg_I(^{187}{\rm Re})=0.0065(3)(5)(2) {\rm~eV}$. 
Here the uncertainties correspond to the uncertainties in Eqs.~(\ref{ABW185}) and (\ref{ABW187}) above.

\section{Conclusion}
In the present paper, we have obtained refined values of the magnetic moments of $^{185}$Re and $^{187}$Re nuclei. For this, we have calculated the shielding constant for the ReO$_4^-$ anion using a combination of the relativistic coupled cluster and relativistic density functional theories. We have studied the influence of the finite nuclear magnetization distribution effect on the shielding constant value. Such effect is usually omitted in molecular calculations. However, according to our study, this effect can be more important than the solvent effect which is often estimated. Updated values of the nuclear magnetic moments resolve the disagreement between theoretical predictions~\cite{shabaev1997ground, beier2000gj, boucard2000relativistic} and experimental values~\cite{Crespo:1998} for the hyperfine splittings of H-like rhenium ions. In addition to the nuclear magnetic moment values, we have also used H-like data for rhenium HFS constants to extract the universal parameter~\cite{Skripnikov:2020e} of the nuclear magnetization distribution. The values of the nuclear magnetic moment and this parameter are necessary ingredients for the theoretical prediction of HFS constants in different rhenium ions and compounds.

\begin{acknowledgments}
We are grateful to V.M. Shabaev for useful discussions. Electronic structure calculations have been carried out using computing resources of the federal collective usage center Complex for Simulation and Data Processing for Mega-science Facilities at National Research Centre ``Kurchatov Institute'', http://ckp.nrcki.ru/, and partly using the computing resources of the quantum chemistry laboratory.

$~~~$Molecular electronic structure calculations performed at NRC ``Kurchatov Institute'' -- PNPI  have been supported by the Russian Science Foundation Grant No. 19-72-10019. Calculations of nucleon wave function  performed at SPbSU were supported by the foundation for the advancement of theoretical physics and mathematics ``BASIS'' grant according to Project No. 21-1-2-47-1.
\end{acknowledgments}


\begin{thebibliography}{85}
\expandafter\ifx\csname natexlab\endcsname\relax\def\natexlab#1{#1}\fi
\expandafter\ifx\csname bibnamefont\endcsname\relax
  \def\bibnamefont#1{#1}\fi
\expandafter\ifx\csname bibfnamefont\endcsname\relax
  \def\bibfnamefont#1{#1}\fi
\expandafter\ifx\csname citenamefont\endcsname\relax
  \def\citenamefont#1{#1}\fi
\expandafter\ifx\csname url\endcsname\relax
  \def\url#1{\texttt{#1}}\fi
\expandafter\ifx\csname urlprefix\endcsname\relax\def\urlprefix{URL }\fi
\providecommand{\bibinfo}[2]{#2}
\providecommand{\eprint}[2][]{\url{#2}}

\bibitem[{\citenamefont{Alarcon et~al.}(2022)\citenamefont{Alarcon, Alexander,
  Anastassopoulos, Aoki, Baartman, Baeßler, Bartoszek, Beck, Bedeschi, Berger
  et~al.}}]{eEDM_snowmass:2022}
\bibinfo{author}{\bibfnamefont{R.}~\bibnamefont{Alarcon}},
  \bibinfo{author}{\bibfnamefont{J.}~\bibnamefont{Alexander}},
  \bibinfo{author}{\bibfnamefont{V.}~\bibnamefont{Anastassopoulos}},
  \bibinfo{author}{\bibfnamefont{T.}~\bibnamefont{Aoki}},
  \bibinfo{author}{\bibfnamefont{R.}~\bibnamefont{Baartman}},
  \bibinfo{author}{\bibfnamefont{S.}~\bibnamefont{Baeßler}},
  \bibinfo{author}{\bibfnamefont{L.}~\bibnamefont{Bartoszek}},
  \bibinfo{author}{\bibfnamefont{D.~H.} \bibnamefont{Beck}},
  \bibinfo{author}{\bibfnamefont{F.}~\bibnamefont{Bedeschi}},
  \bibinfo{author}{\bibfnamefont{R.}~\bibnamefont{Berger}},
  \bibnamefont{et~al.} (\bibinfo{year}{2022}), \bibinfo{note}{arXiv:2203.08103
  [hep-ph](2022)}.

\bibitem[{\citenamefont{Safronova et~al.}(2018)\citenamefont{Safronova, Budker,
  DeMille, Kimball, Derevianko, and Clark}}]{Safronova:18}
\bibinfo{author}{\bibfnamefont{M.~S.} \bibnamefont{Safronova}},
  \bibinfo{author}{\bibfnamefont{D.}~\bibnamefont{Budker}},
  \bibinfo{author}{\bibfnamefont{D.}~\bibnamefont{DeMille}},
  \bibinfo{author}{\bibfnamefont{D.~F.~J.} \bibnamefont{Kimball}},
  \bibinfo{author}{\bibfnamefont{A.}~\bibnamefont{Derevianko}},
  \bibnamefont{and} \bibinfo{author}{\bibfnamefont{C.~W.} \bibnamefont{Clark}},
  \bibinfo{journal}{Rev.\ Mod.\ Phys.} \textbf{\bibinfo{volume}{90}},
  \bibinfo{pages}{025008} (\bibinfo{year}{2018}).

\bibitem[{\citenamefont{Porsev et~al.}(2009)\citenamefont{Porsev, Beloy, and
  Derevianko}}]{Porsev:2009}
\bibinfo{author}{\bibfnamefont{S.~G.} \bibnamefont{Porsev}},
  \bibinfo{author}{\bibfnamefont{K.}~\bibnamefont{Beloy}}, \bibnamefont{and}
  \bibinfo{author}{\bibfnamefont{A.}~\bibnamefont{Derevianko}},
  \bibinfo{journal}{Phys. Rev. Lett.} \textbf{\bibinfo{volume}{102}},
  \bibinfo{pages}{181601} (\bibinfo{year}{2009}).

\bibitem[{\citenamefont{Ginges et~al.}(2017)\citenamefont{Ginges, Volotka, and
  Fritzsche}}]{ginges2017ground}
\bibinfo{author}{\bibfnamefont{J.~S.~M.} \bibnamefont{Ginges}},
  \bibinfo{author}{\bibfnamefont{A.~V.} \bibnamefont{Volotka}},
  \bibnamefont{and}
  \bibinfo{author}{\bibfnamefont{S.}~\bibnamefont{Fritzsche}},
  \bibinfo{journal}{Phys.\ Rev.\ A} \textbf{\bibinfo{volume}{96}},
  \bibinfo{pages}{062502} (\bibinfo{year}{2017}).

\bibitem[{\citenamefont{Fleig and Skripnikov}(2020)}]{Skripnikov:2020b}
\bibinfo{author}{\bibfnamefont{T.}~\bibnamefont{Fleig}} \bibnamefont{and}
  \bibinfo{author}{\bibfnamefont{L.~V.} \bibnamefont{Skripnikov}},
  \bibinfo{journal}{Symmetry} \textbf{\bibinfo{volume}{12}},
  \bibinfo{pages}{498} (\bibinfo{year}{2020}).

\bibitem[{\citenamefont{Ginges and Flambaum}(2004)}]{GFreview}
\bibinfo{author}{\bibfnamefont{J.~S.~M.} \bibnamefont{Ginges}}
  \bibnamefont{and} \bibinfo{author}{\bibfnamefont{V.~V.}
  \bibnamefont{Flambaum}}, \bibinfo{journal}{Phys.\ Rep.}
  \textbf{\bibinfo{volume}{397}}, \bibinfo{pages}{63} (\bibinfo{year}{2004}).

\bibitem[{\citenamefont{Kozlov and Labzowsky}(1995)}]{KL95}
\bibinfo{author}{\bibfnamefont{M.}~\bibnamefont{Kozlov}} \bibnamefont{and}
  \bibinfo{author}{\bibfnamefont{L.}~\bibnamefont{Labzowsky}},
  \bibinfo{journal}{J.\ Phys.\ B} \textbf{\bibinfo{volume}{28}},
  \bibinfo{pages}{1933} (\bibinfo{year}{1995}).

\bibitem[{\citenamefont{Quiney et~al.}(1998)\citenamefont{Quiney, Skaane, and
  Grant}}]{Quiney:98}
\bibinfo{author}{\bibfnamefont{H.~M.} \bibnamefont{Quiney}},
  \bibinfo{author}{\bibfnamefont{H.}~\bibnamefont{Skaane}}, \bibnamefont{and}
  \bibinfo{author}{\bibfnamefont{I.~P.} \bibnamefont{Grant}},
  \bibinfo{journal}{J.\ Phys.\ B} \textbf{\bibinfo{volume}{31}},
  \bibinfo{pages}{L85} (\bibinfo{year}{1998}).

\bibitem[{\citenamefont{Titov et~al.}(2006)\citenamefont{Titov, Mosyagin,
  Petrov, Isaev, and DeMille}}]{Titov:06amin}
\bibinfo{author}{\bibfnamefont{A.~V.} \bibnamefont{Titov}},
  \bibinfo{author}{\bibfnamefont{N.~S.} \bibnamefont{Mosyagin}},
  \bibinfo{author}{\bibfnamefont{A.~N.} \bibnamefont{Petrov}},
  \bibinfo{author}{\bibfnamefont{T.~A.} \bibnamefont{Isaev}}, \bibnamefont{and}
  \bibinfo{author}{\bibfnamefont{D.~P.} \bibnamefont{DeMille}},
  \bibinfo{journal}{Progr.\ Theor.\ Chem.\ Phys.}
  \textbf{\bibinfo{volume}{15}}, \bibinfo{pages}{253} (\bibinfo{year}{2006}).

\bibitem[{\citenamefont{Skripnikov and
  Titov}(2015{\natexlab{a}})}]{Skripnikov:15b}
\bibinfo{author}{\bibfnamefont{L.~V.} \bibnamefont{Skripnikov}}
  \bibnamefont{and} \bibinfo{author}{\bibfnamefont{A.~V.} \bibnamefont{Titov}},
  \bibinfo{journal}{Phys. Rev. A} \textbf{\bibinfo{volume}{91}},
  \bibinfo{pages}{042504} (\bibinfo{year}{2015}{\natexlab{a}}).

\bibitem[{\citenamefont{Skripnikov and
  Titov}(2015{\natexlab{b}})}]{Skripnikov:15a}
\bibinfo{author}{\bibfnamefont{L.~V.} \bibnamefont{Skripnikov}}
  \bibnamefont{and} \bibinfo{author}{\bibfnamefont{A.~V.} \bibnamefont{Titov}},
  \bibinfo{journal}{J.\ Chem.\ Phys.} \textbf{\bibinfo{volume}{142}},
  \bibinfo{eid}{024301} (\bibinfo{year}{2015}{\natexlab{b}}).

\bibitem[{\citenamefont{Sunaga et~al.}(2016)\citenamefont{Sunaga, Abe, Hada,
  and Das}}]{Sunaga:16}
\bibinfo{author}{\bibfnamefont{A.}~\bibnamefont{Sunaga}},
  \bibinfo{author}{\bibfnamefont{M.}~\bibnamefont{Abe}},
  \bibinfo{author}{\bibfnamefont{M.}~\bibnamefont{Hada}}, \bibnamefont{and}
  \bibinfo{author}{\bibfnamefont{B.~P.} \bibnamefont{Das}},
  \bibinfo{journal}{Phys.\ Rev.\ A} \textbf{\bibinfo{volume}{93}},
  \bibinfo{pages}{042507} (\bibinfo{year}{2016}).

\bibitem[{\citenamefont{Fleig}(2017)}]{Fleig:17}
\bibinfo{author}{\bibfnamefont{T.}~\bibnamefont{Fleig}},
  \bibinfo{journal}{Phys.\ Rev.\ A} \textbf{\bibinfo{volume}{96}},
  \bibinfo{pages}{040502(R)} (\bibinfo{year}{2017}).

\bibitem[{\citenamefont{Haase et~al.}(2020)\citenamefont{Haase, Eliav,
  Ilia\v{s}, and Borschevsky}}]{Borschevsky:2020}
\bibinfo{author}{\bibfnamefont{P.~A.~B.} \bibnamefont{Haase}},
  \bibinfo{author}{\bibfnamefont{E.}~\bibnamefont{Eliav}},
  \bibinfo{author}{\bibfnamefont{M.}~\bibnamefont{Ilia\v{s}}},
  \bibnamefont{and}
  \bibinfo{author}{\bibfnamefont{A.}~\bibnamefont{Borschevsky}},
  \bibinfo{journal}{J. Phys. Chem. A} \textbf{\bibinfo{volume}{124}},
  \bibinfo{pages}{3157} (\bibinfo{year}{2020}).

\bibitem[{\citenamefont{Skripnikov}(2020)}]{Skripnikov:2020e}
\bibinfo{author}{\bibfnamefont{L.~V.} \bibnamefont{Skripnikov}},
  \bibinfo{journal}{J.\ Chem.\ Phys.} \textbf{\bibinfo{volume}{153}},
  \bibinfo{pages}{114114} (\bibinfo{year}{2020}).

\bibitem[{\citenamefont{Skripnikov
  et~al.}(2017{\natexlab{a}})\citenamefont{Skripnikov, Titov, and
  Flambaum}}]{Skripnikov:17b}
\bibinfo{author}{\bibfnamefont{L.~V.} \bibnamefont{Skripnikov}},
  \bibinfo{author}{\bibfnamefont{A.~V.} \bibnamefont{Titov}}, \bibnamefont{and}
  \bibinfo{author}{\bibfnamefont{V.~V.} \bibnamefont{Flambaum}},
  \bibinfo{journal}{Phys.\ Rev.\ A} \textbf{\bibinfo{volume}{95}},
  \bibinfo{pages}{022512} (\bibinfo{year}{2017}{\natexlab{a}}).

\bibitem[{\citenamefont{Skripnikov et~al.}(2014)\citenamefont{Skripnikov,
  Kudashov, Petrov, and Titov}}]{Skripnikov:14c}
\bibinfo{author}{\bibfnamefont{L.~V.} \bibnamefont{Skripnikov}},
  \bibinfo{author}{\bibfnamefont{A.~D.} \bibnamefont{Kudashov}},
  \bibinfo{author}{\bibfnamefont{A.~N.} \bibnamefont{Petrov}},
  \bibnamefont{and} \bibinfo{author}{\bibfnamefont{A.~V.} \bibnamefont{Titov}},
  \bibinfo{journal}{Phys.\ Rev.\ A} \textbf{\bibinfo{volume}{90}},
  \bibinfo{pages}{064501} (\bibinfo{year}{2014}).

\bibitem[{\citenamefont{Skripnikov}(2017)}]{Skripnikov:17c}
\bibinfo{author}{\bibfnamefont{L.~V.} \bibnamefont{Skripnikov}},
  \bibinfo{journal}{J.\ Chem.\ Phys.} \textbf{\bibinfo{volume}{147}},
  \bibinfo{pages}{021101} (\bibinfo{year}{2017}).

\bibitem[{\citenamefont{Skripnikov et~al.}(2015)\citenamefont{Skripnikov,
  Petrov, Mosyagin, Titov, and Flambaum}}]{Skripnikov:15c}
\bibinfo{author}{\bibfnamefont{L.~V.} \bibnamefont{Skripnikov}},
  \bibinfo{author}{\bibfnamefont{A.~N.} \bibnamefont{Petrov}},
  \bibinfo{author}{\bibfnamefont{N.~S.} \bibnamefont{Mosyagin}},
  \bibinfo{author}{\bibfnamefont{A.~V.} \bibnamefont{Titov}}, \bibnamefont{and}
  \bibinfo{author}{\bibfnamefont{V.~V.} \bibnamefont{Flambaum}},
  \bibinfo{journal}{Phys.\ Rev.\ A} \textbf{\bibinfo{volume}{92}},
  \bibinfo{pages}{012521} (\bibinfo{year}{2015}).

\bibitem[{\citenamefont{Skripnikov
  et~al.}(2017{\natexlab{b}})\citenamefont{Skripnikov, Maison, and
  Mosyagin}}]{Skripnikov:17a}
\bibinfo{author}{\bibfnamefont{L.~V.} \bibnamefont{Skripnikov}},
  \bibinfo{author}{\bibfnamefont{D.~E.} \bibnamefont{Maison}},
  \bibnamefont{and} \bibinfo{author}{\bibfnamefont{N.~S.}
  \bibnamefont{Mosyagin}}, \bibinfo{journal}{Phys.\ Rev.\ A}
  \textbf{\bibinfo{volume}{95}}, \bibinfo{pages}{022507}
  (\bibinfo{year}{2017}{\natexlab{b}}).

\bibitem[{\citenamefont{Persson}(1998)}]{Persson1998}
\bibinfo{author}{\bibfnamefont{J.}~\bibnamefont{Persson}},
  \bibinfo{journal}{Eur.\ Phys.\ J.\ A} \textbf{\bibinfo{volume}{2}},
  \bibinfo{pages}{3} (\bibinfo{year}{1998}).

\bibitem[{\citenamefont{Cheal and Flanagan}(2010)}]{0954-3899-37-11-113101}
\bibinfo{author}{\bibfnamefont{B.}~\bibnamefont{Cheal}} \bibnamefont{and}
  \bibinfo{author}{\bibfnamefont{K.~T.} \bibnamefont{Flanagan}},
  \bibinfo{journal}{Journal of Physics G: Nuclear and Particle Physics}
  \textbf{\bibinfo{volume}{37}}, \bibinfo{pages}{113101}
  (\bibinfo{year}{2010}).

\bibitem[{\citenamefont{Schmidt et~al.}(2018)\citenamefont{Schmidt, Billowes,
  Bissell, Blaum, Ruiz, Heylen, Malbrunot-Ettenauer, Neyens,
  N{\"o}rtersh{\"a}user, Plunien et~al.}}]{Schmidt:2018}
\bibinfo{author}{\bibfnamefont{S.}~\bibnamefont{Schmidt}},
  \bibinfo{author}{\bibfnamefont{J.}~\bibnamefont{Billowes}},
  \bibinfo{author}{\bibfnamefont{M.~L.} \bibnamefont{Bissell}},
  \bibinfo{author}{\bibfnamefont{K.}~\bibnamefont{Blaum}},
  \bibinfo{author}{\bibfnamefont{R.~F.~G.} \bibnamefont{Ruiz}},
  \bibinfo{author}{\bibfnamefont{H.}~\bibnamefont{Heylen}},
  \bibinfo{author}{\bibfnamefont{S.}~\bibnamefont{Malbrunot-Ettenauer}},
  \bibinfo{author}{\bibfnamefont{G.}~\bibnamefont{Neyens}},
  \bibinfo{author}{\bibfnamefont{W.}~\bibnamefont{N{\"o}rtersh{\"a}user}},
  \bibinfo{author}{\bibfnamefont{G.}~\bibnamefont{Plunien}},
  \bibnamefont{et~al.}, \bibinfo{journal}{Phys.\ Lett.\ B}
  \textbf{\bibinfo{volume}{779}}, \bibinfo{pages}{324 } (\bibinfo{year}{2018}).

\bibitem[{\citenamefont{Barzakh et~al.}(2012)\citenamefont{Barzakh, Batist,
  Fedorov, Ivanov, Mezilev, Molkanov, Moroz, Orlov, Panteleev, and
  Volkov}}]{barzakh2012hyperfine}
\bibinfo{author}{\bibfnamefont{A.~E.} \bibnamefont{Barzakh}},
  \bibinfo{author}{\bibfnamefont{L.~K.} \bibnamefont{Batist}},
  \bibinfo{author}{\bibfnamefont{D.~V.} \bibnamefont{Fedorov}},
  \bibinfo{author}{\bibfnamefont{V.~S.} \bibnamefont{Ivanov}},
  \bibinfo{author}{\bibfnamefont{K.~A.} \bibnamefont{Mezilev}},
  \bibinfo{author}{\bibfnamefont{P.~L.} \bibnamefont{Molkanov}},
  \bibinfo{author}{\bibfnamefont{F.~V.} \bibnamefont{Moroz}},
  \bibinfo{author}{\bibfnamefont{S.~Y.} \bibnamefont{Orlov}},
  \bibinfo{author}{\bibfnamefont{V.~N.} \bibnamefont{Panteleev}},
  \bibnamefont{and} \bibinfo{author}{\bibfnamefont{Y.~M.}
  \bibnamefont{Volkov}}, \bibinfo{journal}{Phys.\ Rev.\ C}
  \textbf{\bibinfo{volume}{86}}, \bibinfo{pages}{014311}
  (\bibinfo{year}{2012}).

\bibitem[{\citenamefont{Prosnyak et~al.}(2020)\citenamefont{Prosnyak, Maison,
  and Skripnikov}}]{Prosnyak:2020}
\bibinfo{author}{\bibfnamefont{S.~D.} \bibnamefont{Prosnyak}},
  \bibinfo{author}{\bibfnamefont{D.~E.} \bibnamefont{Maison}},
  \bibnamefont{and} \bibinfo{author}{\bibfnamefont{L.~V.}
  \bibnamefont{Skripnikov}}, \bibinfo{journal}{J.\ Chem.\ Phys.}
  \textbf{\bibinfo{volume}{152}}, \bibinfo{pages}{044301}
  (\bibinfo{year}{2020}).

\bibitem[{\citenamefont{Roberts and Ginges}(2020)}]{Ginges:2020}
\bibinfo{author}{\bibfnamefont{B.~M.} \bibnamefont{Roberts}} \bibnamefont{and}
  \bibinfo{author}{\bibfnamefont{J.~S.~M.} \bibnamefont{Ginges}},
  \bibinfo{journal}{Phys. Rev. Lett.} \textbf{\bibinfo{volume}{125}},
  \bibinfo{pages}{063002} (\bibinfo{year}{2020}).

\bibitem[{\citenamefont{Barzakh et~al.}(2020)\citenamefont{Barzakh, Atanasov,
  Andreyev, Al~Monthery, Althubiti, Andel, Antalic, Blaum, Cocolios, Cubiss
  et~al.}}]{Barzakh:2020}
\bibinfo{author}{\bibfnamefont{A.~E.} \bibnamefont{Barzakh}},
  \bibinfo{author}{\bibfnamefont{D.}~\bibnamefont{Atanasov}},
  \bibinfo{author}{\bibfnamefont{A.~N.} \bibnamefont{Andreyev}},
  \bibinfo{author}{\bibfnamefont{M.}~\bibnamefont{Al~Monthery}},
  \bibinfo{author}{\bibfnamefont{N.~A.} \bibnamefont{Althubiti}},
  \bibinfo{author}{\bibfnamefont{B.}~\bibnamefont{Andel}},
  \bibinfo{author}{\bibfnamefont{S.}~\bibnamefont{Antalic}},
  \bibinfo{author}{\bibfnamefont{K.}~\bibnamefont{Blaum}},
  \bibinfo{author}{\bibfnamefont{T.~E.} \bibnamefont{Cocolios}},
  \bibinfo{author}{\bibfnamefont{J.~G.} \bibnamefont{Cubiss}},
  \bibnamefont{et~al.}, \bibinfo{journal}{Phys. Rev. C}
  \textbf{\bibinfo{volume}{101}}, \bibinfo{pages}{034308}
  (\bibinfo{year}{2020}).

\bibitem[{\citenamefont{Prosnyak and Skripnikov}(2021)}]{Prosnyak:2021}
\bibinfo{author}{\bibfnamefont{S.~D.} \bibnamefont{Prosnyak}} \bibnamefont{and}
  \bibinfo{author}{\bibfnamefont{L.~V.} \bibnamefont{Skripnikov}},
  \bibinfo{journal}{Phys. Rev. C} \textbf{\bibinfo{volume}{103}},
  \bibinfo{pages}{034314} (\bibinfo{year}{2021}).

\bibitem[{\citenamefont{Shabaev et~al.}(2001)\citenamefont{Shabaev, Artemyev,
  Yerokhin, Zherebtsov, and Soff}}]{Shabaev:01a}
\bibinfo{author}{\bibfnamefont{V.~M.} \bibnamefont{Shabaev}},
  \bibinfo{author}{\bibfnamefont{A.~N.} \bibnamefont{Artemyev}},
  \bibinfo{author}{\bibfnamefont{V.~A.} \bibnamefont{Yerokhin}},
  \bibinfo{author}{\bibfnamefont{O.~M.} \bibnamefont{Zherebtsov}},
  \bibnamefont{and} \bibinfo{author}{\bibfnamefont{G.}~\bibnamefont{Soff}},
  \bibinfo{journal}{Phys.\ Rev.\ Lett.} \textbf{\bibinfo{volume}{86}},
  \bibinfo{pages}{3959} (\bibinfo{year}{2001}).

\bibitem[{\citenamefont{Werth et~al.}(2001)\citenamefont{Werth, H{\"a}ffner,
  Hermanspahn, Kluge, Quint, and Verd{\'u}}}]{werth2001g}
\bibinfo{author}{\bibfnamefont{G.}~\bibnamefont{Werth}},
  \bibinfo{author}{\bibfnamefont{H.}~\bibnamefont{H{\"a}ffner}},
  \bibinfo{author}{\bibfnamefont{N.}~\bibnamefont{Hermanspahn}},
  \bibinfo{author}{\bibfnamefont{H.-J.} \bibnamefont{Kluge}},
  \bibinfo{author}{\bibfnamefont{W.}~\bibnamefont{Quint}}, \bibnamefont{and}
  \bibinfo{author}{\bibfnamefont{J.}~\bibnamefont{Verd{\'u}}}, in
  \emph{\bibinfo{booktitle}{The Hydrogen Atom}} (\bibinfo{publisher}{Springer},
  \bibinfo{year}{2001}), pp. \bibinfo{pages}{204--220}.

\bibitem[{\citenamefont{Quint et~al.}(2008)\citenamefont{Quint, Moskovkhin,
  Shabaev, and Vogel}}]{Quint:2008}
\bibinfo{author}{\bibfnamefont{W.}~\bibnamefont{Quint}},
  \bibinfo{author}{\bibfnamefont{D.~L.} \bibnamefont{Moskovkhin}},
  \bibinfo{author}{\bibfnamefont{V.~M.} \bibnamefont{Shabaev}},
  \bibnamefont{and} \bibinfo{author}{\bibfnamefont{M.}~\bibnamefont{Vogel}},
  \bibinfo{journal}{Phys.\ Rev.\ A} \textbf{\bibinfo{volume}{78}},
  \bibinfo{pages}{032517} (\bibinfo{year}{2008}).

\bibitem[{\citenamefont{Volchkova et~al.}(2017)\citenamefont{Volchkova,
  Varentsova, Zubova, Agababaev, Glazov, Volotka, Shabaev, and
  Plunien}}]{Volchkova:2017}
\bibinfo{author}{\bibfnamefont{A.}~\bibnamefont{Volchkova}},
  \bibinfo{author}{\bibfnamefont{A.}~\bibnamefont{Varentsova}},
  \bibinfo{author}{\bibfnamefont{N.}~\bibnamefont{Zubova}},
  \bibinfo{author}{\bibfnamefont{V.}~\bibnamefont{Agababaev}},
  \bibinfo{author}{\bibfnamefont{D.}~\bibnamefont{Glazov}},
  \bibinfo{author}{\bibfnamefont{A.}~\bibnamefont{Volotka}},
  \bibinfo{author}{\bibfnamefont{V.}~\bibnamefont{Shabaev}}, \bibnamefont{and}
  \bibinfo{author}{\bibfnamefont{G.}~\bibnamefont{Plunien}},
  \bibinfo{journal}{Nuclear Instruments and Methods in Physics Research Section
  B: Beam Interactions with Materials and Atoms}
  \textbf{\bibinfo{volume}{408}}, \bibinfo{pages}{89} (\bibinfo{year}{2017}),
  ISSN \bibinfo{issn}{0168-583X}, \bibinfo{note}{proceedings of the 18th
  International Conference on the Physics of Highly Charged Ions (HCI-2016),
  Kielce, Poland, 11-16 September 2016}.

\bibitem[{\citenamefont{Skripnikov et~al.}(2018)\citenamefont{Skripnikov,
  Schmidt, Ullmann, Geppert, Kraus, Kresse, N\"ortersh\"auser, Privalov,
  Scheibe, Shabaev et~al.}}]{Skripnikov:18a}
\bibinfo{author}{\bibfnamefont{L.~V.} \bibnamefont{Skripnikov}},
  \bibinfo{author}{\bibfnamefont{S.}~\bibnamefont{Schmidt}},
  \bibinfo{author}{\bibfnamefont{J.}~\bibnamefont{Ullmann}},
  \bibinfo{author}{\bibfnamefont{C.}~\bibnamefont{Geppert}},
  \bibinfo{author}{\bibfnamefont{F.}~\bibnamefont{Kraus}},
  \bibinfo{author}{\bibfnamefont{B.}~\bibnamefont{Kresse}},
  \bibinfo{author}{\bibfnamefont{W.}~\bibnamefont{N\"ortersh\"auser}},
  \bibinfo{author}{\bibfnamefont{A.~F.} \bibnamefont{Privalov}},
  \bibinfo{author}{\bibfnamefont{B.}~\bibnamefont{Scheibe}},
  \bibinfo{author}{\bibfnamefont{V.~M.} \bibnamefont{Shabaev}},
  \bibnamefont{et~al.}, \bibinfo{journal}{Phys.\ Rev.\ Lett.}
  \textbf{\bibinfo{volume}{120}}, \bibinfo{pages}{093001}
  (\bibinfo{year}{2018}).

\bibitem[{\citenamefont{Antu\v{s}ek and Repisky}(2020)}]{Antusek:20}
\bibinfo{author}{\bibfnamefont{A.}~\bibnamefont{Antu\v{s}ek}} \bibnamefont{and}
  \bibinfo{author}{\bibfnamefont{M.}~\bibnamefont{Repisky}},
  \bibinfo{journal}{Phys. Chem. Chem. Phys.} \textbf{\bibinfo{volume}{22}},
  \bibinfo{pages}{7065} (\bibinfo{year}{2020}).

\bibitem[{\citenamefont{Antu\ifmmode~\check{s}\else \v{s}\fi{}ek
  et~al.}(2018)\citenamefont{Antu\ifmmode~\check{s}\else \v{s}\fi{}ek, Repisky,
  Jaszu\ifmmode~\acute{n}\else \'{n}\fi{}ski, Jackowski, Makulski, and
  Misiak}}]{Antusek:18}
\bibinfo{author}{\bibfnamefont{A.}~\bibnamefont{Antu\ifmmode~\check{s}\else
  \v{s}\fi{}ek}}, \bibinfo{author}{\bibfnamefont{M.}~\bibnamefont{Repisky}},
  \bibinfo{author}{\bibfnamefont{M.}~\bibnamefont{Jaszu\ifmmode~\acute{n}\else
  \'{n}\fi{}ski}}, \bibinfo{author}{\bibfnamefont{K.}~\bibnamefont{Jackowski}},
  \bibinfo{author}{\bibfnamefont{W.}~\bibnamefont{Makulski}}, \bibnamefont{and}
  \bibinfo{author}{\bibfnamefont{M.}~\bibnamefont{Misiak}},
  \bibinfo{journal}{Phys. Rev. A} \textbf{\bibinfo{volume}{98}},
  \bibinfo{pages}{052509} (\bibinfo{year}{2018}).

\bibitem[{\citenamefont{Alder and Yu}(1951)}]{Alder:1951}
\bibinfo{author}{\bibfnamefont{F.}~\bibnamefont{Alder}} \bibnamefont{and}
  \bibinfo{author}{\bibfnamefont{F.~C.} \bibnamefont{Yu}},
  \bibinfo{journal}{Phys. Rev.} \textbf{\bibinfo{volume}{82}},
  \bibinfo{pages}{105} (\bibinfo{year}{1951}).

\bibitem[{\citenamefont{Stone}(2014)}]{stone2014}
\bibinfo{author}{\bibfnamefont{N.}~\bibnamefont{Stone}},
  \bibinfo{journal}{Table of nuclear magnetic dipole and electric quadrupole
  moments, INDC(NDS)--0658, International Atomic Energy Agency (IAEA)}
  (\bibinfo{year}{2014}).

\bibitem[{\citenamefont{Feiock and Johnson}(1968)}]{Johnson:68}
\bibinfo{author}{\bibfnamefont{F.~D.} \bibnamefont{Feiock}} \bibnamefont{and}
  \bibinfo{author}{\bibfnamefont{W.~R.} \bibnamefont{Johnson}},
  \bibinfo{journal}{Phys.\ Rev.\ Lett.} \textbf{\bibinfo{volume}{21}},
  \bibinfo{pages}{785} (\bibinfo{year}{1968}).

\bibitem[{\citenamefont{Feiock and Johnson}(1969)}]{Feiock:1969}
\bibinfo{author}{\bibfnamefont{F.~D.} \bibnamefont{Feiock}} \bibnamefont{and}
  \bibinfo{author}{\bibfnamefont{W.~R.} \bibnamefont{Johnson}},
  \bibinfo{journal}{Phys.\ Rev.} \textbf{\bibinfo{volume}{187}},
  \bibinfo{pages}{39} (\bibinfo{year}{1969}).

\bibitem[{\citenamefont{Kolb et~al.}(1982)\citenamefont{Kolb, Johnson, and
  Shorer}}]{Kolb:1982}
\bibinfo{author}{\bibfnamefont{D.}~\bibnamefont{Kolb}},
  \bibinfo{author}{\bibfnamefont{W.~R.} \bibnamefont{Johnson}},
  \bibnamefont{and} \bibinfo{author}{\bibfnamefont{P.}~\bibnamefont{Shorer}},
  \bibinfo{journal}{Phys.\ Rev.\ A} \textbf{\bibinfo{volume}{26}},
  \bibinfo{pages}{19} (\bibinfo{year}{1982}).

\bibitem[{\citenamefont{Crespo L\'opez-Urrutia
  et~al.}(1998)\citenamefont{Crespo L\'opez-Urrutia, Beiersdorfer, Widmann,
  Birkett, M\aa{}rtensson-Pendrill, and Gustavsson}}]{Crespo:1998}
\bibinfo{author}{\bibfnamefont{J.~R.} \bibnamefont{Crespo L\'opez-Urrutia}},
  \bibinfo{author}{\bibfnamefont{P.}~\bibnamefont{Beiersdorfer}},
  \bibinfo{author}{\bibfnamefont{K.}~\bibnamefont{Widmann}},
  \bibinfo{author}{\bibfnamefont{B.~B.} \bibnamefont{Birkett}},
  \bibinfo{author}{\bibfnamefont{A.-M.} \bibnamefont{M\aa{}rtensson-Pendrill}},
  \bibnamefont{and} \bibinfo{author}{\bibfnamefont{M.~G.~H.}
  \bibnamefont{Gustavsson}}, \bibinfo{journal}{Phys.\ Rev.\ A}
  \textbf{\bibinfo{volume}{57}}, \bibinfo{pages}{879} (\bibinfo{year}{1998}).

\bibitem[{\citenamefont{Dyall and F{\ae}gri~Jr}(2007)}]{DyallFaegri2007}
\bibinfo{author}{\bibfnamefont{K.~G.} \bibnamefont{Dyall}} \bibnamefont{and}
  \bibinfo{author}{\bibfnamefont{K.}~\bibnamefont{F{\ae}gri~Jr}},
  \emph{\bibinfo{title}{Introduction to relativistic quantum chemistry}}
  (\bibinfo{publisher}{Oxford University Press}, \bibinfo{year}{2007}).

\bibitem[{\citenamefont{Bohr and Weisskopf}(1950)}]{bohr1950influence}
\bibinfo{author}{\bibfnamefont{A.}~\bibnamefont{Bohr}} \bibnamefont{and}
  \bibinfo{author}{\bibfnamefont{V.~F.} \bibnamefont{Weisskopf}},
  \bibinfo{journal}{Phys.\ Rev.} \textbf{\bibinfo{volume}{77}},
  \bibinfo{pages}{94} (\bibinfo{year}{1950}).

\bibitem[{\citenamefont{Bohr}(1951)}]{bohr1951bohr}
\bibinfo{author}{\bibfnamefont{A.}~\bibnamefont{Bohr}},
  \bibinfo{journal}{Phys.\ Rev.} \textbf{\bibinfo{volume}{81}},
  \bibinfo{pages}{134} (\bibinfo{year}{1951}).

\bibitem[{\citenamefont{Sliv}(1951)}]{sliv1951uchet}
\bibinfo{author}{\bibfnamefont{L.}~\bibnamefont{Sliv}}, \bibinfo{journal}{Zh.
  Eksp. Teor. Fiz.} \textbf{\bibinfo{volume}{21}}, \bibinfo{pages}{770}
  (\bibinfo{year}{1951}).

\bibitem[{\citenamefont{Zherebtsov and Shabaev}(2000)}]{Zherebtsov:2000}
\bibinfo{author}{\bibfnamefont{O.~M.} \bibnamefont{Zherebtsov}}
  \bibnamefont{and} \bibinfo{author}{\bibfnamefont{V.~M.}
  \bibnamefont{Shabaev}}, \bibinfo{journal}{Can. J. Phys.}
  \textbf{\bibinfo{volume}{78}}, \bibinfo{pages}{701} (\bibinfo{year}{2000}).

\bibitem[{\citenamefont{Shabaev et~al.}(1997)\citenamefont{Shabaev, Tomaselli,
  K{\"u}hl, Artemyev, and Yerokhin}}]{shabaev1997ground}
\bibinfo{author}{\bibfnamefont{V.~M.} \bibnamefont{Shabaev}},
  \bibinfo{author}{\bibfnamefont{M.}~\bibnamefont{Tomaselli}},
  \bibinfo{author}{\bibfnamefont{T.}~\bibnamefont{K{\"u}hl}},
  \bibinfo{author}{\bibfnamefont{A.~N.} \bibnamefont{Artemyev}},
  \bibnamefont{and} \bibinfo{author}{\bibfnamefont{V.~A.}
  \bibnamefont{Yerokhin}}, \bibinfo{journal}{Phys.\ Rev.\ A}
  \textbf{\bibinfo{volume}{56}}, \bibinfo{pages}{252} (\bibinfo{year}{1997}).

\bibitem[{\citenamefont{Tupitsyn et~al.}(2002)\citenamefont{Tupitsyn, Loginov,
  and Shabaev}}]{Tupitsyn:02}
\bibinfo{author}{\bibfnamefont{I.~I.} \bibnamefont{Tupitsyn}},
  \bibinfo{author}{\bibfnamefont{A.~V.} \bibnamefont{Loginov}},
  \bibnamefont{and} \bibinfo{author}{\bibfnamefont{V.~M.}
  \bibnamefont{Shabaev}}, \bibinfo{journal}{Optics and Spectroscopy}
  \textbf{\bibinfo{volume}{93}}, \bibinfo{pages}{357} (\bibinfo{year}{2002}).

\bibitem[{\citenamefont{Volotka et~al.}(2008)\citenamefont{Volotka, Glazov,
  Tupitsyn, Oreshkina, Plunien, and Shabaev}}]{Volotka:2008}
\bibinfo{author}{\bibfnamefont{A.~V.} \bibnamefont{Volotka}},
  \bibinfo{author}{\bibfnamefont{D.~A.} \bibnamefont{Glazov}},
  \bibinfo{author}{\bibfnamefont{I.~I.} \bibnamefont{Tupitsyn}},
  \bibinfo{author}{\bibfnamefont{N.~S.} \bibnamefont{Oreshkina}},
  \bibinfo{author}{\bibfnamefont{G.}~\bibnamefont{Plunien}}, \bibnamefont{and}
  \bibinfo{author}{\bibfnamefont{V.~M.} \bibnamefont{Shabaev}},
  \bibinfo{journal}{Phys. Rev. A} \textbf{\bibinfo{volume}{78}},
  \bibinfo{pages}{062507} (\bibinfo{year}{2008}).

\bibitem[{\citenamefont{Malkin et~al.}(2011)\citenamefont{Malkin, Repiský,
  Komorovský, Mach, Malkina, and Malkin}}]{Malkin:2011}
\bibinfo{author}{\bibfnamefont{E.}~\bibnamefont{Malkin}},
  \bibinfo{author}{\bibfnamefont{M.}~\bibnamefont{Repiský}},
  \bibinfo{author}{\bibfnamefont{S.}~\bibnamefont{Komorovský}},
  \bibinfo{author}{\bibfnamefont{P.}~\bibnamefont{Mach}},
  \bibinfo{author}{\bibfnamefont{O.~L.} \bibnamefont{Malkina}},
  \bibnamefont{and} \bibinfo{author}{\bibfnamefont{V.~G.}
  \bibnamefont{Malkin}}, \bibinfo{journal}{J.\ Chem.\ Phys.}
  \textbf{\bibinfo{volume}{134}}, \bibinfo{pages}{044111}
  (\bibinfo{year}{2011}).

\bibitem[{\citenamefont{Roberts and Ginges}(2021)}]{Roberts:2021}
\bibinfo{author}{\bibfnamefont{B.~M.} \bibnamefont{Roberts}} \bibnamefont{and}
  \bibinfo{author}{\bibfnamefont{J.~S.~M.} \bibnamefont{Ginges}},
  \bibinfo{journal}{Phys. Rev. A} \textbf{\bibinfo{volume}{104}},
  \bibinfo{pages}{022823} (\bibinfo{year}{2021}).

\bibitem[{\citenamefont{Sapirstein and Cheng}(2003)}]{Sapirstein:2003}
\bibinfo{author}{\bibfnamefont{J.}~\bibnamefont{Sapirstein}} \bibnamefont{and}
  \bibinfo{author}{\bibfnamefont{K.~T.} \bibnamefont{Cheng}},
  \bibinfo{journal}{Phys.\ Rev.\ A} \textbf{\bibinfo{volume}{67}},
  \bibinfo{pages}{022512} (\bibinfo{year}{2003}).

\bibitem[{\citenamefont{Konovalova et~al.}(2017)\citenamefont{Konovalova,
  Kozlov, Demidov, and Barzakh}}]{konovalova2017calculation}
\bibinfo{author}{\bibfnamefont{E.~A.} \bibnamefont{Konovalova}},
  \bibinfo{author}{\bibfnamefont{M.~G.} \bibnamefont{Kozlov}},
  \bibinfo{author}{\bibfnamefont{Y.~A.} \bibnamefont{Demidov}},
  \bibnamefont{and} \bibinfo{author}{\bibfnamefont{A.~E.}
  \bibnamefont{Barzakh}}, \bibinfo{journal}{Rad. Appl.}
  \textbf{\bibinfo{volume}{2}}, \bibinfo{pages}{181} (\bibinfo{year}{2017}),
  \urlprefix\url{arXiv:1703.10048}.

\bibitem[{\citenamefont{Ginges and Volotka}(2018)}]{Ginges:2018}
\bibinfo{author}{\bibfnamefont{J.}~\bibnamefont{Ginges}} \bibnamefont{and}
  \bibinfo{author}{\bibfnamefont{A.}~\bibnamefont{Volotka}},
  \bibinfo{journal}{Phys.\ Rev.\ A} \textbf{\bibinfo{volume}{98}},
  \bibinfo{pages}{032504} (\bibinfo{year}{2018}).

\bibitem[{\citenamefont{Kozlov et~al.}(2001)\citenamefont{Kozlov, Porsev, and
  Johnson}}]{kozlov2001parity}
\bibinfo{author}{\bibfnamefont{M.~G.} \bibnamefont{Kozlov}},
  \bibinfo{author}{\bibfnamefont{S.~G.} \bibnamefont{Porsev}},
  \bibnamefont{and} \bibinfo{author}{\bibfnamefont{W.~R.}
  \bibnamefont{Johnson}}, \bibinfo{journal}{Phys.\ Rev.\ A}
  \textbf{\bibinfo{volume}{64}}, \bibinfo{pages}{052107}
  (\bibinfo{year}{2001}).

\bibitem[{\citenamefont{Shabaev}(1994)}]{shabaev1994hyperfine}
\bibinfo{author}{\bibfnamefont{V.~M.} \bibnamefont{Shabaev}},
  \bibinfo{journal}{J.\ Phys.\ B} \textbf{\bibinfo{volume}{27}},
  \bibinfo{pages}{5825} (\bibinfo{year}{1994}).

\bibitem[{\citenamefont{Aucar et~al.}(1999)\citenamefont{Aucar, Saue, Visscher,
  and Jensen}}]{Aucar:99}
\bibinfo{author}{\bibfnamefont{G.~A.} \bibnamefont{Aucar}},
  \bibinfo{author}{\bibfnamefont{T.}~\bibnamefont{Saue}},
  \bibinfo{author}{\bibfnamefont{L.}~\bibnamefont{Visscher}}, \bibnamefont{and}
  \bibinfo{author}{\bibfnamefont{H.~J.~A.} \bibnamefont{Jensen}},
  \bibinfo{journal}{J.\ Chem.\ Phys.} \textbf{\bibinfo{volume}{110}},
  \bibinfo{pages}{6208} (\bibinfo{year}{1999}).

\bibitem[{\citenamefont{Olejniczak et~al.}(2012)\citenamefont{Olejniczak, Bast,
  Saue, and Pecul}}]{Olejniczak:12}
\bibinfo{author}{\bibfnamefont{M.}~\bibnamefont{Olejniczak}},
  \bibinfo{author}{\bibfnamefont{R.}~\bibnamefont{Bast}},
  \bibinfo{author}{\bibfnamefont{T.}~\bibnamefont{Saue}}, \bibnamefont{and}
  \bibinfo{author}{\bibfnamefont{M.}~\bibnamefont{Pecul}},
  \bibinfo{journal}{J.\ Chem.\ Phys.} \textbf{\bibinfo{volume}{136}},
  \bibinfo{pages}{014108} (\bibinfo{year}{2012}).

\bibitem[{\citenamefont{Ilias et~al.}(2013)\citenamefont{Ilias, Jensen, Bast,
  and Saue}}]{Ilias:13}
\bibinfo{author}{\bibfnamefont{M.}~\bibnamefont{Ilias}},
  \bibinfo{author}{\bibfnamefont{H.~J.~A.} \bibnamefont{Jensen}},
  \bibinfo{author}{\bibfnamefont{R.}~\bibnamefont{Bast}}, \bibnamefont{and}
  \bibinfo{author}{\bibfnamefont{T.}~\bibnamefont{Saue}},
  \bibinfo{journal}{Mol. Phys.} \textbf{\bibinfo{volume}{111}},
  \bibinfo{pages}{1373} (\bibinfo{year}{2013}).

\bibitem[{DIR()}]{DIRAC15}
\bibinfo{note}{DIRAC, a relativistic ab initio electronic structure program,
  Release DIRAC15 (2015), written by R. Bast, T. Saue, L. Visscher, and H. J.
  Aa. Jensen, with contributions from V. Bakken, K. G. Dyall, S. Dubillard, U.
  Ekstroem, E. Eliav, T. Enevoldsen, E. Fasshauer, T. Fleig, O. Fossgaard, A.
  S. P. Gomes, T. Helgaker, J. Henriksson, M. Ilias, Ch. R. Jacob, S. Knecht,
  S. Komorovsky, O. Kullie, J. K. Laerdahl, C. V. Larsen, Y. S. Lee, H. S.
  Nataraj, M. K. Nayak, P. Norman, G. Olejniczak, J. Olsen, Y. C. Park, J. K.
  Pedersen, M. Pernpointner, R. Di Remigio, K. Ruud, P. Salek, B.
  Schimmelpfennig, J. Sikkema, A. J. Thorvaldsen, J. Thyssen, J. van Stralen,
  S. Villaume, O. Visser, T. Winther, and S. Yamamoto (see
  http://www.diracprogram.org).}

\bibitem[{\citenamefont{Saue et~al.}(2020)\citenamefont{Saue, Gomes, Jensen,
  Visscher, Aucar, {Di Remigio}, Dyall, Eliav, Fasshauer, Fleig
  et~al.}}]{Saue:2020}
\bibinfo{author}{\bibfnamefont{R.}~\bibnamefont{Saue}, \bibfnamefont{T.~Bast}},
  \bibinfo{author}{\bibfnamefont{A.~S.~P.} \bibnamefont{Gomes}},
  \bibinfo{author}{\bibfnamefont{H.~J.~A.} \bibnamefont{Jensen}},
  \bibinfo{author}{\bibfnamefont{L.}~\bibnamefont{Visscher}},
  \bibinfo{author}{\bibfnamefont{I.~A.} \bibnamefont{Aucar}},
  \bibinfo{author}{\bibfnamefont{R.}~\bibnamefont{{Di Remigio}}},
  \bibinfo{author}{\bibfnamefont{K.~G.} \bibnamefont{Dyall}},
  \bibinfo{author}{\bibfnamefont{E.}~\bibnamefont{Eliav}},
  \bibinfo{author}{\bibfnamefont{E.}~\bibnamefont{Fasshauer}},
  \bibinfo{author}{\bibfnamefont{T.}~\bibnamefont{Fleig}},
  \bibnamefont{et~al.}, \bibinfo{journal}{J. Chem. Phys.}
  \textbf{\bibinfo{volume}{152}}, \bibinfo{pages}{204104}
  (\bibinfo{year}{2020}).

\bibitem[{\citenamefont{Dyall}(2007)}]{Dyall:07}
\bibinfo{author}{\bibfnamefont{K.~G.} \bibnamefont{Dyall}},
  \bibinfo{journal}{Theor. Chem. Acc.} \textbf{\bibinfo{volume}{117}},
  \bibinfo{pages}{491} (\bibinfo{year}{2007}).

\bibitem[{\citenamefont{Dyall}(2012)}]{Dyall:12}
\bibinfo{author}{\bibfnamefont{K.~G.} \bibnamefont{Dyall}},
  \bibinfo{journal}{Theor. Chem. Acc.} \textbf{\bibinfo{volume}{131}},
  \bibinfo{pages}{1217} (\bibinfo{year}{2012}).

\bibitem[{\citenamefont{{Dunning, Jr}}(1989)}]{Dunning:89}
\bibinfo{author}{\bibfnamefont{T.~H.} \bibnamefont{{Dunning, Jr}}},
  \bibinfo{journal}{J.\ Chem.\ Phys.} \textbf{\bibinfo{volume}{90}},
  \bibinfo{pages}{1007} (\bibinfo{year}{1989}).

\bibitem[{\citenamefont{Kendall et~al.}(1992)\citenamefont{Kendall, {Dunning,
  Jr}, and Harrison}}]{Kendall:92}
\bibinfo{author}{\bibfnamefont{R.~A.} \bibnamefont{Kendall}},
  \bibinfo{author}{\bibfnamefont{T.~H.} \bibnamefont{{Dunning, Jr}}},
  \bibnamefont{and} \bibinfo{author}{\bibfnamefont{R.~J.}
  \bibnamefont{Harrison}}, \bibinfo{journal}{J.\ Chem.\ Phys.}
  \textbf{\bibinfo{volume}{96}}, \bibinfo{pages}{6796} (\bibinfo{year}{1992}).

\bibitem[{\citenamefont{Adamo and Barone}(1999)}]{pbe0}
\bibinfo{author}{\bibfnamefont{C.}~\bibnamefont{Adamo}} \bibnamefont{and}
  \bibinfo{author}{\bibfnamefont{V.}~\bibnamefont{Barone}},
  \bibinfo{journal}{J.\ Chem.\ Phys.} \textbf{\bibinfo{volume}{110}},
  \bibinfo{pages}{6158} (\bibinfo{year}{1999}).

\bibitem[{MRC()}]{MRCC2020}
\bibinfo{note}{M. K\'{a}llay, P. R. Nagy, D. Mester, Z. Rolik, G. Samu, J.
  Csontos, J. Cs\'{o}ka, P. B. Szab\'{o}, L. Gyevi-Nagy, B. H\'{e}gely, I.
  Ladj\'{a}nszki, L. Szegedy, B. Lad\'{o}czki, K. Petrov, M. Farkas, P. D.
  Mezei, and \'{a}. Ganyecz: The {\sc mrcc} program system: Accurate quantum
  chemistry from water to proteins, J. Chem. Phys. 152, 074107 (2020).”
  “{\sc mrcc}, a quantum chemical program suite written by M. K\'{a}llay, P.
  R. Nagy, D. Mester, Z. Rolik, G. Samu, J. Csontos, J. Cs\'{o}ka, P. B.
  Szab\'{o}, L. Gyevi-Nagy, B. H\'{e}gely, I. Ladj\'{a}nszki, L. Szegedy, B.
  Lad\'{o}czki, K. Petrov, M. Farkas, P. D. Mezei, and \'{a}. Ganyecz. See
  www.mrcc.hu.}

\bibitem[{\citenamefont{Visscher and Dyall}(1997)}]{Visscher:1997}
\bibinfo{author}{\bibfnamefont{L.}~\bibnamefont{Visscher}} \bibnamefont{and}
  \bibinfo{author}{\bibfnamefont{K.~G.} \bibnamefont{Dyall}},
  \bibinfo{journal}{Atomic Data and Nuclear Data Tables}
  \textbf{\bibinfo{volume}{67}}, \bibinfo{pages}{207} (\bibinfo{year}{1997}).

\bibitem[{\citenamefont{Fella et~al.}(2020)\citenamefont{Fella, Skripnikov,
  N\"ortersh\"auser, Buchner, Deubner, Kraus, Privalov, Shabaev, and
  Vogel}}]{Skripnikov:2020a}
\bibinfo{author}{\bibfnamefont{V.}~\bibnamefont{Fella}},
  \bibinfo{author}{\bibfnamefont{L.~V.} \bibnamefont{Skripnikov}},
  \bibinfo{author}{\bibfnamefont{W.}~\bibnamefont{N\"ortersh\"auser}},
  \bibinfo{author}{\bibfnamefont{M.~R.} \bibnamefont{Buchner}},
  \bibinfo{author}{\bibfnamefont{H.~L.} \bibnamefont{Deubner}},
  \bibinfo{author}{\bibfnamefont{F.}~\bibnamefont{Kraus}},
  \bibinfo{author}{\bibfnamefont{A.~F.} \bibnamefont{Privalov}},
  \bibinfo{author}{\bibfnamefont{V.~M.} \bibnamefont{Shabaev}},
  \bibnamefont{and} \bibinfo{author}{\bibfnamefont{M.}~\bibnamefont{Vogel}},
  \bibinfo{journal}{Phys. Rev. Research} \textbf{\bibinfo{volume}{2}},
  \bibinfo{pages}{013368} (\bibinfo{year}{2020}).

\bibitem[{\citenamefont{Visscher et~al.}(1996)\citenamefont{Visscher, Lee, and
  Dyall}}]{Visscher:96a}
\bibinfo{author}{\bibfnamefont{L.}~\bibnamefont{Visscher}},
  \bibinfo{author}{\bibfnamefont{T.~J.} \bibnamefont{Lee}}, \bibnamefont{and}
  \bibinfo{author}{\bibfnamefont{K.~G.} \bibnamefont{Dyall}},
  \bibinfo{journal}{J.\ Chem.\ Phys.} \textbf{\bibinfo{volume}{105}},
  \bibinfo{pages}{8769} (\bibinfo{year}{1996}).

\bibitem[{\citenamefont{Yerokhin et~al.}(2011)\citenamefont{Yerokhin, Pachucki,
  Harman, and Keitel}}]{Yerokhin:2011}
\bibinfo{author}{\bibfnamefont{V.~A.} \bibnamefont{Yerokhin}},
  \bibinfo{author}{\bibfnamefont{K.}~\bibnamefont{Pachucki}},
  \bibinfo{author}{\bibfnamefont{Z.}~\bibnamefont{Harman}}, \bibnamefont{and}
  \bibinfo{author}{\bibfnamefont{C.~H.} \bibnamefont{Keitel}},
  \bibinfo{journal}{Phys. Rev. Lett.} \textbf{\bibinfo{volume}{107}},
  \bibinfo{pages}{043004} (\bibinfo{year}{2011}).

\bibitem[{\citenamefont{Yerokhin et~al.}(2012)\citenamefont{Yerokhin, Pachucki,
  Harman, and Keitel}}]{Yerokhin:2012}
\bibinfo{author}{\bibfnamefont{V.~A.} \bibnamefont{Yerokhin}},
  \bibinfo{author}{\bibfnamefont{K.}~\bibnamefont{Pachucki}},
  \bibinfo{author}{\bibfnamefont{Z.}~\bibnamefont{Harman}}, \bibnamefont{and}
  \bibinfo{author}{\bibfnamefont{C.~H.} \bibnamefont{Keitel}},
  \bibinfo{journal}{Phys. Rev. A} \textbf{\bibinfo{volume}{85}},
  \bibinfo{pages}{022512} (\bibinfo{year}{2012}).

\bibitem[{\citenamefont{Becke}(1993)}]{b3lyp}
\bibinfo{author}{\bibfnamefont{A.}~\bibnamefont{Becke}}, \bibinfo{journal}{J.\
  Chem.\ Phys.} \textbf{\bibinfo{volume}{98}}, \bibinfo{pages}{5648}
  (\bibinfo{year}{1993}).

\bibitem[{\citenamefont{Kozio\l{} et~al.}(2019)\citenamefont{Kozio\l{}, Aucar,
  and Aucar}}]{Koziol:2019}
\bibinfo{author}{\bibfnamefont{K.}~\bibnamefont{Kozio\l{}}},
  \bibinfo{author}{\bibfnamefont{I.~A.} \bibnamefont{Aucar}}, \bibnamefont{and}
  \bibinfo{author}{\bibfnamefont{G.~A.} \bibnamefont{Aucar}},
  \bibinfo{journal}{J.\ Chem.\ Phys.} \textbf{\bibinfo{volume}{150}},
  \bibinfo{pages}{184301} (\bibinfo{year}{2019}).

\bibitem[{\citenamefont{Gustavsson and
  M\aa{}rtensson-Pendrill}(1998)}]{Gustavsson:1998}
\bibinfo{author}{\bibfnamefont{M.~G.~H.} \bibnamefont{Gustavsson}}
  \bibnamefont{and} \bibinfo{author}{\bibfnamefont{A.-M.}
  \bibnamefont{M\aa{}rtensson-Pendrill}}, \bibinfo{journal}{Phys.\ Rev.\ A}
  \textbf{\bibinfo{volume}{58}}, \bibinfo{pages}{3611} (\bibinfo{year}{1998}).

\bibitem[{\citenamefont{Lutz}(1967)}]{Lutz:1967}
\bibinfo{author}{\bibfnamefont{O.}~\bibnamefont{Lutz}}, \bibinfo{journal}{Phys.
  Lett. A} \textbf{\bibinfo{volume}{25}}, \bibinfo{pages}{440}
  (\bibinfo{year}{1967}), ISSN \bibinfo{issn}{0375-9601}.

\bibitem[{\citenamefont{Raghavan}(1989)}]{Raghavan:89}
\bibinfo{author}{\bibfnamefont{P.}~\bibnamefont{Raghavan}},
  \bibinfo{journal}{Atomic Data and Nuclear Data Tables}
  \textbf{\bibinfo{volume}{42}}, \bibinfo{pages}{189 } (\bibinfo{year}{1989}).

\bibitem[{\citenamefont{Crespo López-Urrutia et~al.}(1997)\citenamefont{Crespo
  López-Urrutia, Beiersdorfer, Savin, and Widmann}}]{CrespoBeiersdorfer97}
\bibinfo{author}{\bibfnamefont{J.~R.} \bibnamefont{Crespo López-Urrutia}},
  \bibinfo{author}{\bibfnamefont{P.}~\bibnamefont{Beiersdorfer}},
  \bibinfo{author}{\bibfnamefont{D.~W.} \bibnamefont{Savin}}, \bibnamefont{and}
  \bibinfo{author}{\bibfnamefont{K.}~\bibnamefont{Widmann}},
  \bibinfo{journal}{AIP Conference Proceedings} \textbf{\bibinfo{volume}{392}},
  \bibinfo{pages}{87} (\bibinfo{year}{1997}).

\bibitem[{Note1()}]{Note1}
\bibinfo{note}{Hyperfine splittings given in Ref.~\cite {Crespo:1998}
  are equal to $3A$ for the case of considered H-like $^{185}$Re and $^{187}$Re
  ions in the ground electronic states.}

\bibitem[{\citenamefont{Artemyev et~al.}(2001)\citenamefont{Artemyev, Shabaev,
  Plunien, Soff, and Yerokhin}}]{artemyev2001vacuum}
\bibinfo{author}{\bibfnamefont{A.}~\bibnamefont{Artemyev}},
  \bibinfo{author}{\bibfnamefont{V.}~\bibnamefont{Shabaev}},
  \bibinfo{author}{\bibfnamefont{G.}~\bibnamefont{Plunien}},
  \bibinfo{author}{\bibfnamefont{G.}~\bibnamefont{Soff}}, \bibnamefont{and}
  \bibinfo{author}{\bibfnamefont{V.}~\bibnamefont{Yerokhin}},
  \bibinfo{journal}{Phys.\ Rev.\ A} \textbf{\bibinfo{volume}{63}},
  \bibinfo{pages}{062504} (\bibinfo{year}{2001}).

\bibitem[{Note2()}]{Note2}
\bibinfo{note}{The uncertainty of $A^{(0)}/g_II$ is estimated as the
  difference between the values of $A^{(0)}/g_II$ calculated within the
  Fermi~\cite {shabaev1997ground} and uniform charge distribution models~\cite
  {shabaev1994hyperfine}. $A^{(0)}/g_II=0.2926(3)$~eV was calculated in~\cite
  {shabaev1997ground} for $^{185}{\protect \rm Re}$. Constant
  for $^{187}{\protect \rm Re}$ was obtained here
  using the constant for $^{185}{\protect \rm Re}$ and the known
  dependence~\cite {shabaev1997ground} of the charge distribution correction on
  the charge radius. Difference of the $A^{(0)}/g_II$ values for
  $^{185}{\protect \rm Re}$ and $^{187}{\protect \rm Re}$ was found to be negligible.}

\bibitem[{Note3()}]{Note3}
\bibinfo{note}{So, the estimated uncertainty of the used single-particle
  model of the nuclear magnetization distribution is about 30\%. Similar
  uncertainty $\delta _{\protect \rm FMD}$ (in \%) has been found for the
  contribution of the nuclear magnetization distribution to the shielding
  constant in section~\ref {sectUncert}}.

\bibitem[{\citenamefont{Beier}(2000)}]{beier2000gj}
\bibinfo{author}{\bibfnamefont{T.}~\bibnamefont{Beier}},
  \bibinfo{journal}{Physics Reports} \textbf{\bibinfo{volume}{339}},
  \bibinfo{pages}{79} (\bibinfo{year}{2000}).

\bibitem[{\citenamefont{Boucard and
  Indelicato}(2000)}]{boucard2000relativistic}
\bibinfo{author}{\bibfnamefont{S.}~\bibnamefont{Boucard}} \bibnamefont{and}
  \bibinfo{author}{\bibfnamefont{P.}~\bibnamefont{Indelicato}},
  \bibinfo{journal}{The European Physical Journal D}
  \textbf{\bibinfo{volume}{8}}, \bibinfo{pages}{59} (\bibinfo{year}{2000}).
  
\bibitem[{\citenamefont{Roberts et~al.}(2022)\citenamefont{Roberts, Ranclaud,
  and Ginges}}]{Ginges:2022}
\bibinfo{author}{\bibfnamefont{B.~M.} \bibnamefont{Roberts}},
  \bibinfo{author}{\bibfnamefont{P.~G.} \bibnamefont{Ranclaud}},
  \bibnamefont{and} \bibinfo{author}{\bibfnamefont{J.~S.~M.}
  \bibnamefont{Ginges}}, \bibinfo{journal}{Phys. Rev. A}
  \textbf{\bibinfo{volume}{105}}, \bibinfo{pages}{052802}  

\end{thebibliography}

\end{document}